\documentclass[useAMS,usenatbib]{mn2e}

\usepackage{epsf}
\usepackage{amsmath}
\usepackage{times}

\newcommand{\hmpc}{h^{-1}{\rm Mpc}}

\newcommand{\asf}{\alpha_{\rm sf}}

\title[SFR-M$_*$ and an evolving IMF]{The galaxy stellar mass--star formation rate relation: Evidence for an evolving stellar initial mass function?}

\author[R. Dav\'e]{Romeel Dav\'e\\Astronomy Department, University of Arizona, Tucson, AZ 85721}

\begin{document}

\pubyear{2007}

\maketitle

\label{firstpage}

 \begin{abstract}
The evolution of the galaxy stellar mass--star formation rate relationship
($M_*-$SFR) provides key constraints on the stellar mass assembly
histories of galaxies.  For star-forming galaxies, $M_*-$SFR is observed
to be fairly tight with a slope close to unity from $z\sim 0\rightarrow
2$, and it evolves downwards roughly independently of $M_*$.  Simulations
of galaxy formation reproduce these trends, broadly independent of
modeling details, owing to the generic dominance of smooth and steady
cold accretion in these systems.  In contrast, the observed amplitude
of the $M_*-$SFR relation evolves markedly differently than in models,
indicating either that stellar mass assembly is poorly understood or
that observations have been misinterpreted.  Stated in terms of a star
formation activity parameter $\asf\equiv (M_*/SFR)/(t_{\rm Hubble}-{\rm
1 Gyr})$, models predict a constant $\asf\sim 1$ out to redshifts
$z\sim 4+$, while the observed $M_*-$SFR relation indicates that $\asf$
increases by $\sim\times 3$ from $z\sim 2$ until today.  The low $\asf$
(i.e. rapid star formation) at high-$z$ not only conflicts with models,
but is also difficult to reconcile with other observations of high-$z$
galaxies, such as the small scatter in $M_*-$SFR, the slow evolution of
star forming galaxies at $z\sim 2-4$, and the modest passive fractions
in mass-selected samples.  Systematic biases could significantly
affect measurements of $M_*$ and SFR, but detailed considerations
suggest that none are obvious candidates to reconcile the discrepancy.
A speculative solution is considered in which the stellar initial
mass function (IMF) evolves towards more high-mass star formation
at earlier epochs.  Following Larson (1998), a model is investigated
in which the characteristic mass $\hat{M}$ where the IMF turns over
increases with redshift.  Population synthesis models are used to show
that the observed and predicted $M_*-$SFR evolution may be brought into
general agreement if $\hat{M}=0.5(1+z)^{2} M_\odot$ out to $z\sim 2$.
Such IMF evolution matches recent observations of cosmic stellar mass growth,
and the resulting $z=0$ cumulative IMF is similar to the ``paunchy" IMF
favored by \citet{far07} to reconcile the observed cosmic star formation
history with present-day fossil light measures.
\end{abstract}

 
\section{Introduction}

How galaxies build up their stellar mass is a central question in galaxy
formation.  Within the broadly successful cold dark matter scenario, gas
is believed to accrete gravitationally into growing dark matter halos,
and then radiative processes enable the gas to decouple from non-baryonic
matter and eventually settle into a star-forming disk~\citep[e.g.][]{mo98}.
However, many complications arise when testing this scenario against
observations, as the stellar assembly of galaxies involves a host of
other processes including star formation, kinetic and thermal feedback
from various sources, and merger-induced activity, all of which are
poorly understood in comparison to halo assembly.

A key insight into gas accretion processes is the recent recognition
of the importance of ``cold mode" accretion~\citep{kat03,bir03,ker05},
where gas infalling from the intergalactic medium (IGM) does not pass
through an accretion shock on its way to forming stars.  Simulations and
analytic models show that cold mode dominates global accretion at $z\ga
2$~\citep{ker05}, and dominates accretion in all halos with masses $\la
10^{12}$M$_\odot$~\citep{bir03,ker05}.

The central features of cold mode accretion are that it is (1) {\it
rapid} (2) {\it smooth}, and (3) {\it steady}.  It is rapid because it
is limited by the free-fall time and not the cooling time; it is smooth
because most of the accretion occurs in small lumps and not through
major mergers~\citep{mur01,ker05,guo07}; and it is steady because it
is governed by the gravitational potential of the slowly growing halo.
A consequence of cold mode accretion is that the star formation rate
is a fairly steady function of time~\citep[e.g.][]{fin06,fin07},
which results in a tight relationship between the stellar mass $M_*$
and the star formation rate SFR.  Hence simulations of star-forming
galaxies generically predict a tight relationship with $M_*\propto$SFR
that evolves slowly with redshift.  In detail, a slope slightly below
unity occurs owing to the growth of hot halos around higher-mass
galaxies that retards accretion.

The stellar mass--star formation rate ($M_*-$SFR) relation
for star forming galaxies has now been observed out to $z\sim
2$, thanks to improving multiwavelength surveys.  As expected
from models, the relationship is seen to be fairly tight from
$z\sim 0-2$, with a slope just below unity and a scatter of $\la
0.3$~dex~\citep{noe07a,elb07,dad07}.  It also evolves downwards
in amplitude to lower redshift in lock step fashion, suggesting a
quiescent global quenching mechanism such as a lowering of the ambient
cosmic density, again as expected in models.  The range of data used
to quantify this evolution is impressive: \citet{noe07a} used the
AEGIS multi-wavelength survey in the Extended Groth Strip to quantify
the $M_*$-SFR relation from $z\sim 1\rightarrow 0$; \citet{elb07} used
GOODS at $z\sim 0.8-1.2$ and SDSS spectra at $z\sim 0$; and \citet{dad07}
used star-forming BzK-selected galaxies with {\it Spitzer}, X-ray, and
radio follow-up to study the relation at $z\sim 1.4-2.5$.  In each case,
a careful accounting was done of both direct UV and re-radiated infrared
photons to measure the total galaxy SFR, paying particular attention to
contamination from active galactic nuclei (AGN) emission (discussed in
\S\ref{sec:syst}).  Multiwavelength data covering the rest-optical were
employed to accurately estimate $M_*$.  This broad agreement in $M_*-$SFR
slope, scatter, and qualitative evolution between model predictions and
these data lends support to the idea that cold mode accretion dominates
in these galaxies.

Yet all is not well for theory.  Closer inspection reveals that the
{\it amplitude} of the $M_*-$SFR relation evolves with time in a way
that is inconsistent with model expectations.  This disagreement is
fairly generic, as shown in \S\ref{sec:comparison}, arising in both
hydrodynamic simulations and semi-analytic models, and is fairly insensitive to
feedback implementation.  The sense of the disagreement is that going
to higher redshifts, the observed SFRs are higher, and/or stellar masses
lower, than predicted in current models.

The purpose of this paper is to understand the implications of the
observed $M_*-$SFR relation for our theoretical view of how galaxies
accumulate stellar mass.  In \S\ref{sec:tact} it is argued that $M_*-$SFR
amplitude evolution implies a typical galaxy star formation history
that is difficult to reconcile with not only model expectations, but
also other observations of high-$z$ galaxies.  Possible systematic
effects that may bias the estimation of SFR and $M_*$ are considered
in \S\ref{sec:syst}, and it is argued that none of them are obvious
candidates to explain the discrepancy.  Finally a speculative avenue
for reconciliation is considered, namely that the stellar initial mass
function (IMF) has a characteristic mass that evolves with redshift
(\S\ref{sec:imf}).  This model is constrained based on the $M_*$-SFR
relation in \S\ref{sec:imfmodel}. The implications for such an evolving
IMF are discussed in \S\ref{sec:mchar}.  Results are summarized in
\S\ref{sec:summary}.  A $\Lambda$CDM cosmology with $\Omega=0.25$ and
$H_0=70$~km/s/Mpc is assumed throughout for computing cosmic timescales.

\section{$M_*-$SFR: Models vs. observations}\label{sec:comparison}

To begin, the observed $M_*-$SFR relation is compared with model
predictions, in order to understand the origin of the relation.
The primary simulations used are cosmological hydrodynamic
runs with Gadget-2~\citep{spr05} incorporating momentum-driven
outflows, which uniquely match a wide variety of IGM and galaxy
observations~\citep[e.g.][]{opp06,dav06,fin07b}.  The new simulations
include stellar recycling assuming a Chabrier IMF, by returning gas (and
metals) into the ambient reservoir, in addition to the usual heating,
cooling, and star formation, and feedback recipes~\citep{spr03,opp06}.
Outflow parameters are computed based on properties of galaxies identified
with an on-the-fly group finder.  These and other improvements that go
towards making simulations more realistic down to $z=0$ are described in
\citet{opp07}.  Two runs to $z=0$ are considered here, each with 256$^3$
dark matter and an equal number of gas particle, in cubic volumes
of 32 and 64$\hmpc$ (comoving) on a side.  The assumed cosmology is
$\Lambda$CDM with $\Omega_m=0.3$, $\Omega_\Lambda=0.7$, $H_0=69$~km/s/Mpc,
$\Omega_b=0.048$, and $\sigma_8=0.83$.  The resolved galaxy population,
defined here as those with stellar masses greater than 64 gas particle
masses, is limited by stellar masses of $2.5\times 10^9M_\odot$ and
$2\times 10^{10}M_\odot$ in the two volumes~\citep[this is somewhat
more conservative than in][in order to ensure accurate star formation
histories]{fin06}.

To explore the impact of feedback parameters, simulations with two other
outflow models are considered, namely one with no outflows, and one with
the \citet{spr03} ``constant wind" outflow model, taken from the suite
of outflow runs described in \citet{opp06}.  These runs have 32$\hmpc$
box sizes with $2\times 256^3$ particles, and they were evolved only down
to $z=2$.  Their cosmology is slightly different, having $\sigma_8=0.9$
from first-year WMAP parameters, but this is not a large effect at the
epochs considered here.

In hydro simulations, star formation rates and stellar masses
are obtained by summing over particles in galaxies identified
using the group finder SKID (Spline Kernel Interpolative DENMAX;
http://www-hpcc.astro.washington.edu/tools/skid.html); see \citet{ker05}
for details.  Gas particles that have high enough density to form
stars have instantaneous star formation rates computed in Gadget-2;
it is this SFR that is summed to obtain the instantaneous SFR of a
galaxy.  Of course, observed galaxy SFR's and $M_*$'s are measured quite
differently than in simulations, a point that will figure prominently
in \S\ref{sec:syst}.  Note that these hydro simulations do not include
any feedback from active galactic nuclei (AGN), or any process that
explicitly truncates star formation in massive galaxies, hence they do
not reproduce the evolution of passive galaxies.  Here the focus is
on star-forming galaxies only, with the assumption that
galaxies are quickly squelched to become passive and so do not appear
in star-forming selected samples~\citep[e.g.][]{sal07}.  Since all 
galaxies in our simulations would be observationally classified
'as star-forming, no cuts are made to the simulated galaxy
population.

To complement the hydro simulations, results for $M_*-$SFR are taken
from the Millenium simulation plus semi-analytic model (``Millenium SAM")
lightcones of~\citet{kit07}, as fit by \citet{dad07} and \citet{elb07}.
Besides being a completely different modeling approach, the Millenium
SAM also includes AGN feedback that broadly reproduces the observed
passive galaxy population and its evolution.  Additionally, a different
semi-analytic model based on the Millenium simulation by \citet{del07}
was examined, to see if variations in semi-analytic prescriptions
can result in significant differences in $M_*-$SFR.  It was found
that the \citet{del07} model produced a somewhat shallower slope than
the \citet{kit07} model, which e.g. at $z\sim 2$ resulted in a smaller
offset from observations at $M_*\sim 10^{10}M_\odot$, but a larger one at
$M_*\sim 10^{11}M_\odot$.  Since the results are qualitatively similar,
only the \citet{kit07} models are discussed further.  While there are
many more semi-analytic models in the literature, and perhaps even some
that are in better agreement with the $M_*-$SFR relation, it is beyond
the scope of this work to consider them all.

\begin{figure}
\vskip -0.5in
\setlength{\epsfxsize}{0.6\textwidth}
\centerline{\epsfbox{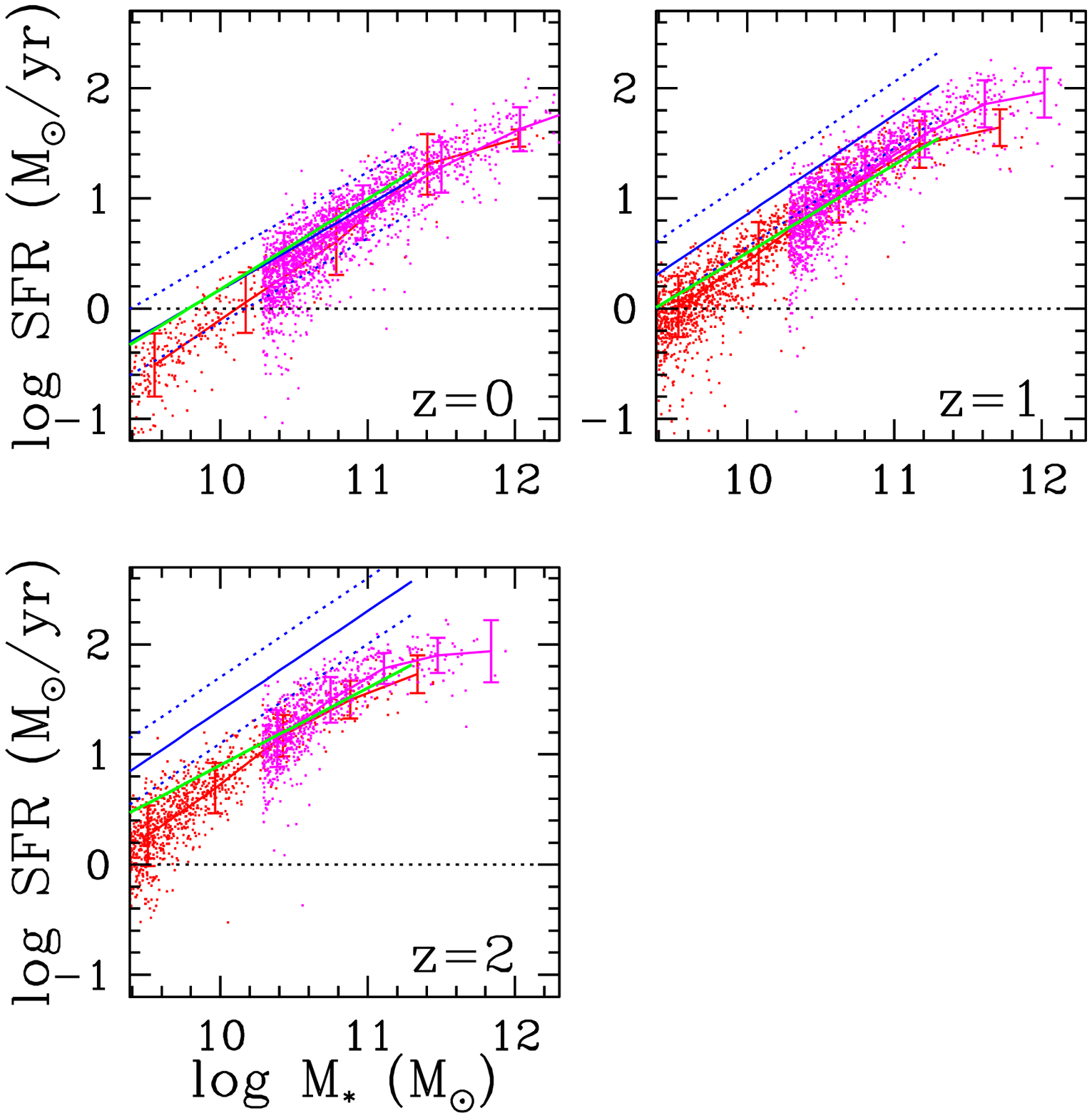}}
\vskip -5.6in
\setlength{\epsfxsize}{0.6\textwidth}
\centerline{\epsfbox{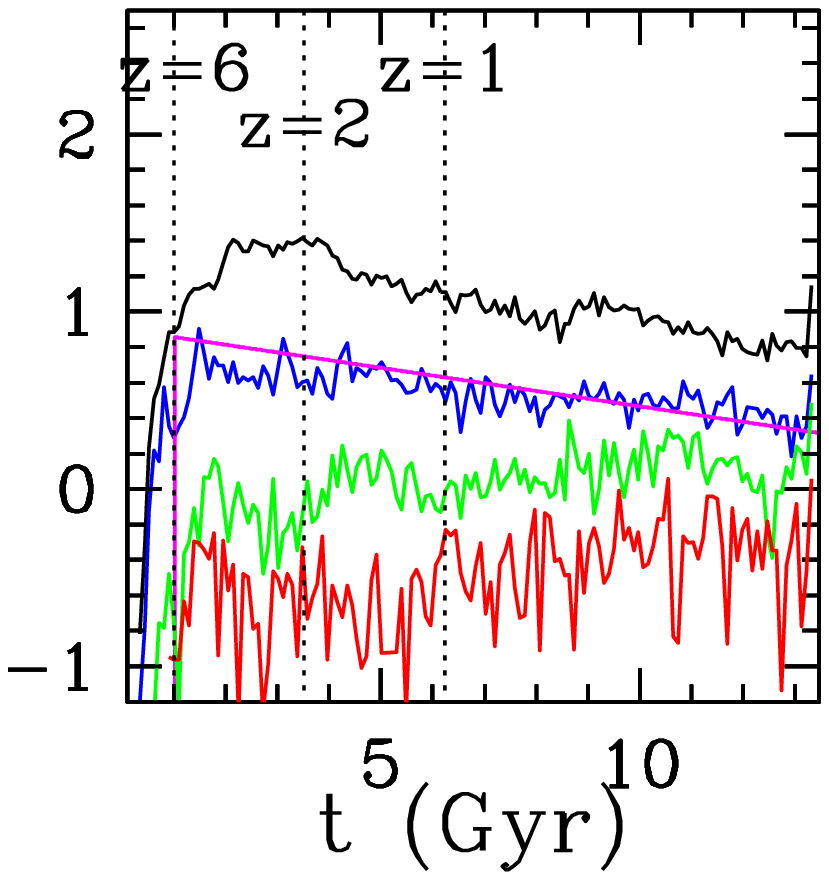}}
\vskip -1.3in
\caption{$M_*-$SFR relation for simulations versus
observations at $z=0,1,2$. Red points show a $32\hmpc$ hydro simulation
box, magenta show $64\hmpc$ box.  A running mean (with standard deviations
shown as vertical bars) is shown
separately for each simulation; agreement in overlap region is excellent.
Green lines show results from the Millenium SAM.  Blue lines show fits
to observations at $z=2$~\citep{dad07}, and $z=0,1$~\citep{elb07}, with
dashed lines indicating $\pm0.3$~dex scatter~\citep{noe07a}.  The models'
slopes and scatter are in good agreement, but the amplitude deviates
from observations at high-$z$.
{\bf Lower right:} Star formation histories in 100~Myr bins for
four star forming galaxies selected from the $32\hmpc$ run closest to
a final $\log M_*=11,10.5,10,9.5$ (top to bottom curves). 
Note the lack of large bursts and the similarity of the star formation history 
shapes, which yields the prediction of a tight $M_*\propto$SFR 
(roughly) relation.  The magenta line shows the form of the SFH assumed 
in PEGASE modeling discussed in \S\ref{sec:imfmodel}.
}
\label{fig:sfrmstar}
\end{figure}

Figure~\ref{fig:sfrmstar} (first three panels) shows the $M_*-$SFR
relation at $z=0,1,2$ in simulations, compared to parameterizations
from observations (blue lines) by \citet{elb07} ($z\approx
0,1$) and \citet{dad07} ($z\approx 2$).  The observed scatter of
0.3~dex~\citep{noe07a,elb07,dad07} is indicated by the dashed blue lines.
It is worth noting that other $z\sim 2$ observations suggest a larger
scatter~\citep{sha05,pap06}, so the observational picture on this is
not entirely settled.  Finally, the green lines show the \citet{kit07}
Millenium SAM results, which generally track the hydro simulation results,
particularly in amplitude which will be the key aspect.

At $z=0$, the relations are generally in agreement, though the observed
slope of $M_*\propto$SFR$^{0.77}$ is somewhat shallower than that
predicted by the hydro simulations ($\approx 0.9$).  The Millenium
SAM produces excellent agreement at $z=0$, having broadly been
tuned to do so.  By $z=1$, the predicted hydro simulation slopes
are similar to that observed by \citet{elb07} though steeper than
seen by \citet{noe07a}, but more importantly the amplitude is low by
$\sim\times 2-3$.  At $z=2$, this trend continues, with a $\sim\times 4-5$
amplitude offset.  The Millenium SAM has a slope that remains near the
$z=0$ value, hence it is somewhat shallower than observed at $z\sim 2$,
but the amplitude evolves slowly much like in the hydro simulations.
The rapid evolution in observed $M_*-$SFR amplitude has been noted
by several authors~\citep{pap06,noe07a}, along with the fact that it
is more rapid than in the Millenium SAM~\citep{elb07,dad07}.  The new
aspect demonstrated here is that the disagreement persists for current
hydrodynamic simulations, and as will be shown in the next section this
is true regardless of the main free parameter in such models, namely
feedback implementation.

The slope and scatter are a direct reflection of the similarity in
the shape of galaxy star formation histories (SFH) among various models and
at various masses.  For a galaxy observed at time $t_0$, neglecting 
mass loss from stellar evolution,
\begin{equation}
M_*(t_0) = \int^{t_0}_0 {{\rm SFR}(t) dt = {{\rm SFR}(t_0) \int^{t_0}_0 \frac{{\rm SFR}(t)}{{\rm SFR}(t_0)}}} dt.
\end{equation}
The integral depends purely on the {\it form} of the SFH up to $t_0$.
So long as the SFH form does not vary significantly with $M_*$, the slope
of $M_*-$SFR will be near unity, and so long as the current SFR is similar
to the recent past-averaged one, the scatter will be small.  As pointed
out by \citet{noe07b}, the observed tight scatter of $\la 0.3$~dex
implies that the bulk of star-forming galaxies cannot be undergoing
large bursts; this will be discussed further in \S\ref{sec:tact}.
This is consistent with a fairly steady mode of galaxy assembly as
generically predicted by cold mode accretion.  Note however that the
reverse conclusion does not necessarily hold, in the sense that a
slope near unity is not uniquely indicative of a steady mode of galaxy
formation; other scenarios \citep[such as the staged galaxy formation
model of][]{noe07b} can be constructed to reproduce a slope near unity,
as discussed in \S\ref{sec:tact}.

Example galaxy SFHs from the $32\hmpc$ momentum-driven
outflow hydro simulation are shown in the lower right panel of
Figure~\ref{fig:sfrmstar}, for a range of $z=0$ stellar masses from
$10^{11}M_\odot-10^{9.5}M_\odot$, in 100~Myr bins.  They are remarkably
self-similar, and generally quite smooth.  The increased variance in
smaller galaxies owes to discreteness effects in spawning star particles
(these SFHs are computed from the history of galaxy star particles
rather than the instantaneous SFR at each time); the true scatter in
instantaneous SFR increases only mildly at small masses, from $\sim
0.2$~dex at $M_*\ga 10^{11}\; M_\odot$ to $\sim 0.3$~dex at  $M_*\ga
10^{9.5}\; M_\odot$, as shown in Figure~\ref{fig:sfrmstar}.

The early simulated SFHs are characterized by a rapid rise in
star formation to $z\sim 6$, corresponding to rapid early halo
growth~\citep{li07}.  Next, there is a period of roughly constant star
formation from $z\sim 6\rightarrow 2$; this will be the key epoch for
the present work.  Finally, at $z\la 2$ the star formation rates decline
modestly, with more massive systems showing greater decline.  This pattern
reflects the changing a balance between the growth of potential wells
being able to attract more matter, and cosmic expansion which lowers the
ambient IGM density and infall rates.  It is qualitatively similar to
the SFH inferred for the Milky Way.  In the models, the fact that the SFH
depends weakly on $M_*$ gives rise to $M_*\propto$SFR, and the fact that
merger-driven starbursts are not a significant growth path for galaxy
stellar mass~\citep{ker05,guo07} gives rise to the small scatter.  In
detail a slope slightly below unity occurs because more massive galaxies
tend to have their accretion curtailed by hot halos~\citep[e.g.][]{ker05},
resulting in ``natural downsizing"~\citep{nei06}.

Note that the SFRs of individual galaxies in simulations do not drop by a
large amount over a Hubble time.  Even for the most massive star forming
galaxies today, the SFR in simulations drops only be a factor of few
since $z=2$, and smaller galaxies show essentially no drop.  Meanwhile,
the amplitude of $M_*$-SFR drops by nearly an order of magnitude in
the models, and even more in the data.  So the amplitude evolution
owes predominantly to the growth of $M_*$ of individual galaxies.
In other words, models indicate that the $M_*-$SFR relation doesn't
evolve downwards in SFR so much as upwards in $M_*$.

Overall, while the slope and scatter of the $M_*-$SFR relation are
in broad agreement, there is a stark and generic discrepancy in the
amplitude evolution between the models and data.  The agreement in
slope and scatter is an encouraging sign that our understanding of gas
accretion processes is reasonably sound.  Examining the discrepancy in
amplitude evolution is the focus of the remainder of this paper.

\section{Evolution of the star formation activity parameter}\label{sec:tact}

\begin{figure}
\vskip -0.5in
\setlength{\epsfxsize}{0.6\textwidth}
\centerline{\epsfbox{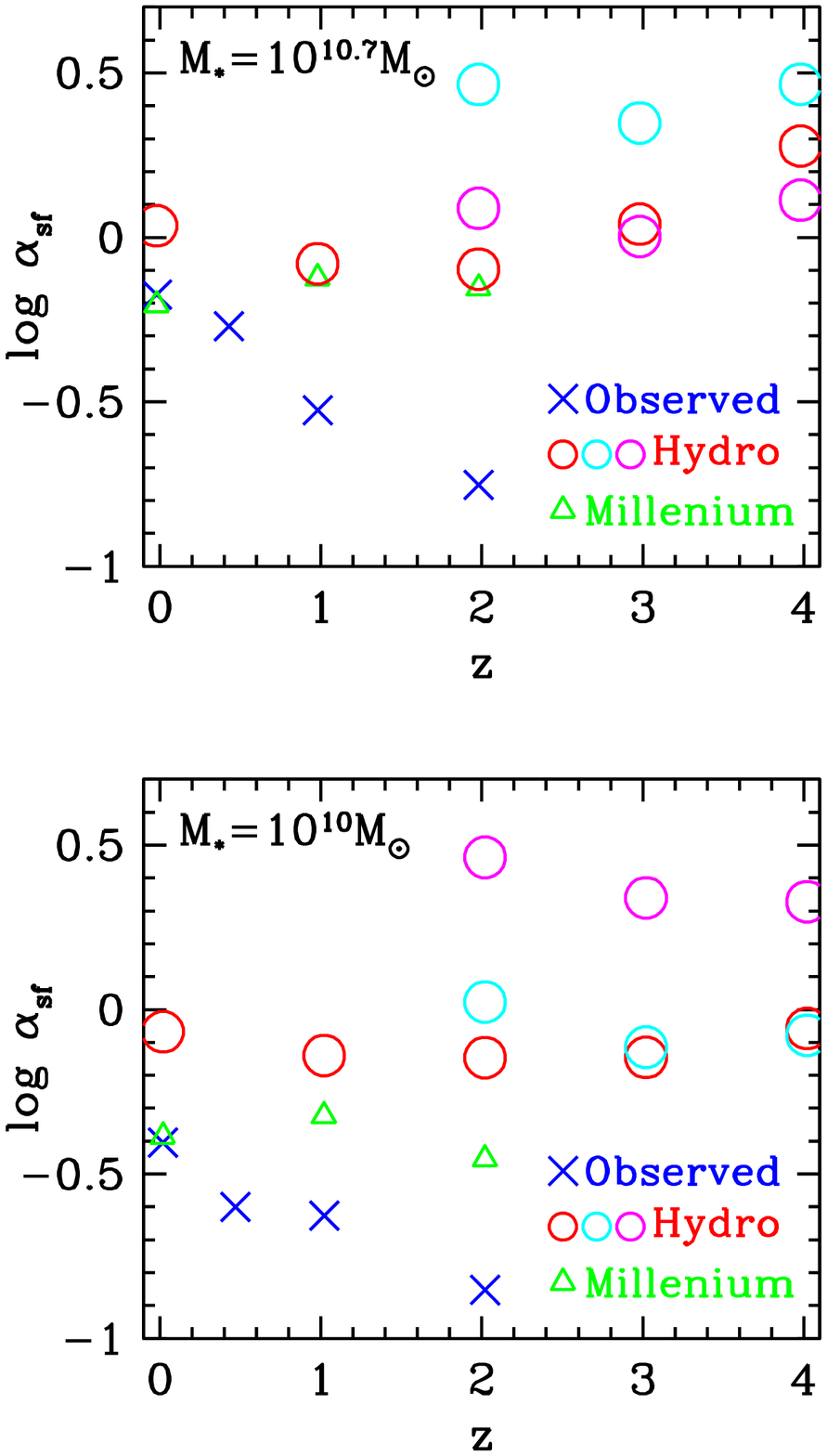}}
\vskip -0.4in
\caption{{\it Top panel:} Star formation activity parameter
$\asf\equiv (M_*/$SFR$)/(t_H-1{\rm Gyr})$ at $z=0-4$, evaluated at
$M_*=10^{10.7}M_\odot$.  Red circles show results from hydro simulations
with a momentum-driven outflow model, cyan circles ($z\geq 2$) show
hydro simulations with no outflows, and magenta circles ($z\geq 2$
show ``constant wind" outflow simulations~\citep[see][for details of
outflow models]{opp06}.  Green triangles show the Millenium SAM results,
and blue crosses show the observations by \citet[$z\approx 0,1$]{elb07},
\citet[$z\approx 0.45$]{noe07a}, and \citet[$z\approx 2$]{dad07}.  In all
cases, regardless of modeling technique or outflow implementation, $\asf$
is predicted to be around unity or larger at all epochs and at all masses.
The observed evolution of $\asf$ out to $z\sim 2$ is markedly discrepant
from model predictions.
{\it Bottom panel:} Same as above for $M_*=10^{10}M_\odot$, showing that
the trends are broadly insensitive to stellar mass.
}
\label{fig:tact}
\end{figure}

The amplitude evolution of $M_*-$SFR can be recast in terms of the
timescale for star formation activity.  Models generally predict that
galaxies have $M_*/$SFR$\sim t_H$, where $t_H$ is the Hubble time,
perhaps reduced by 1~Gyr because few stars were forming at $z\ga 6$,
as seen e.g. in the lower right panel of Figure~\ref{fig:sfrmstar}.

A convenient parameterization is provided by the {\it star formation
activity parameter}\footnote{This is similar but not identical to the
``maturity parameter" of \citet{sco07}.} defined here as the dimensionless
quantity $\asf\equiv (M_*/$SFR$)/(t_H-1{\rm Gyr})$.  Physically, this
may be regarded as the fraction of the Hubble time (minus a Gyr) that a
galaxy needs to have formed stars at its current rate in order to produce
its current stellar mass.  Note that $\asf$ is in general a function of $M_*$.

Figure~\ref{fig:tact} shows $\asf(z)$ for simulations and observations,
top panel showing values at $M_*=10^{10.7}M_\odot$, bottom showing
$M_*=10^{10}M_\odot$.  The values for models have been obtained by
fitting power laws to simulated galaxies' $M_*-$SFR.  The momentum-driven
outflow hydro simulation results are shown as the red circles, while
cyan and magenta circles show results for no winds and the wind model of
\citet{spr03} at $z=2-4$, respectively.  The outflow model has only a
mild effect on $\asf$, because the {\it form} of the SFHs are broadly
similar regardless of outflows, even though the total amount of stars
formed is significantly different~\citep{opp06}.  The most dramatic 
outlier among models is the high-mass bin of the no-wind simulation,
where the rapid gas consumption in large galaxies results in lower current
star formation rates relative to the past-averaged one (i.e. natural
downsizing); this actually makes $\asf$ larger and the discrepancy
with observations worse.  Millenium SAM results are quite similar to
the hydro runs, again showing little evolution in $\asf$, although
the shallower slope results in a somewhat smaller $\asf$ for the lower
mass bin.  Overall, it is remarkable that regardless of methodology or
feedback processes, all models predict $\asf$ to be around unity with
little evolution from $z=4\rightarrow 0$, regardless of stellar mass.
This suggests that $\asf\sim 1$ is a generic consequence of smooth and
steady cold accretion dominating the growth of these galaxies.

In contrast, the observed $M_*-$SFR relation yields a rapidly rising
$\asf=0.19\rightarrow 0.31\rightarrow 0.75$ from $z=2\rightarrow
1\rightarrow 0$ at $10^{10.7}M_\odot$.  Values at $10^{10}M_\odot$
are slightly lower but show similar rapid evolution.  The dramatic
difference between observed and predicted amplitude evolution either
indicates a significant misunderstanding in our basic picture of galaxy
assembly, or else that observations have been misinterpreted in some way.
Given the currently uncertain state of galaxy formation theory, one
might be inclined towards the former, and call into question the entire
picture of cold mode-dominated accretion.  However, it turns out that
the rapidly declining $\asf$ also results in uncomfortable conflicts
with other {\it observations} of high-redshift galaxies.  To see this,
let us discard the idea of cold mode accretion, or any preconceived
theoretical notion, and ask: What possible star formation histories for
these galaxies could yield $\asf$ rising with time as observed?

One scenario is to postulate that observed high-$z$ galaxies are
progressively more dominated by bursts of star formation.  Indeed, some
models have suggested that bursts may comprise a significant portion
of the high-$z$ galaxy population~\citep[e.g.][]{kol99}.  However, as
\citet{noe07a} points out, this seems unlikely given the tightness of
the $M_*$-SFR relation.  They argue that the observed scatter of 0.3~dex
at $z\sim 1$ implies that most star forming galaxies have current star
formation rates $\la \times 2-3$ of their quiescent value, and that
the mean of the relation represents the quiescent ``main sequence"
of galaxy assembly.  Few galaxies between this main sequence and the
passive population are seen, arguing for rapid movement between the
two populations.  At $z\sim 2$, the constraints are even more severe,
as \citet{dad07} measure a $1\sigma$ scatter of 0.29~dex (0.16~dex
interquartile), while the observed $\asf\approx 0.2$ implies they
are bursting at $\sim\times 5$ their quiescent rate.  Since these
observations are able to probe SFRs down to $\sim 10\; M_\odot$/yr,
moderate stellar mass ($M_*\ga 10^{10.5}M_\odot$) galaxies with $\asf\sim
1$ would have been seen if present, but are not.  The only way the burst
scenario would work is if all observed galaxies somehow conspire to be
bursting at approximately the same level, which seems highly contrived.
Hence bursts do not seem to be a viable way to produce low $\asf$.

Another possible scenario is that models simply need to begin forming
stars at a later epoch.  This would effectively lower $\asf$ as defined
here, while still maintaining a fairly constant SFR as required by the
observed scatter.  Theoretically, such a scenario is easier said than
done, as the epoch where galaxy formation begins is not a free parameter
even in semi-analytic models, but rather is governed by halo formation
times that are well-constrained by current cosmological parameters.
Such a scenario would require overwhelmingly strong feedback at
early epochs, far stronger than even the strong feedback present in
e.g. hydro simulations that match observed $z\sim 4-7$ galaxy star
formation rates~\citep{dav06,bou07}, in order for copious dense gas to
be present in early halos yet not form stars.  Still, in the spirit of
discarding any theoretical preconceptions, ``delayed galaxy formation"
appears to be a viable solution.

However, quantitative consideration of delayed galaxy formation
leads to significant problems when confronted by other observations.
At $z=2$, $\asf\approx 0.2$ implies that (constant) star formation
began roughly 0.5~Gyr prior -- which corresponds to $z\approx 2.3$!
Of course it cannot be true that {\it all} galaxies began forming stars
at $z\sim 2.3$, as many star-forming galaxies are seen at significantly
higher redshifts~\citep[e.g.][]{bou07}.  Indeed, both the cosmic star
formation rate \citep[e.g.][]{hop06,far07}, and the rest-UV luminosity
function~\citep{red07} are observed to be fairly constant between $z\sim
4\rightarrow 2$.  So it appears galaxies were forming stars quite
vigorously at $z\gg 2.3$.  Even if galaxies smoothly ramped up their
star formation to values observed at $z\sim 2$, they would still need
to start at $z\approx 2.7$.

If most galaxies cannot have such a low $\asf$, perhaps the observed
systems represent only a small subset of galaxies at $z\sim 2$, and the
rest are simply unseen.  Since current surveys identify both UV-luminous
and red, dusty star forming galaxies, it must be that the unseen galaxies
are passive, or else so extremely extincted that they only show up in
e.g. sub-millimeter surveys.  For the latter case, \citet{red05} found
that the cosmic SFR has a relatively minor contribution from such severely
extincted systems that escape BzK/LBG selection at $z\sim 2$, so they are
not likely to be the dominant growth mode for galaxies.  Moreover, this
actually exacerbates the $M_*-$SFR problem, because it provides a mode to
form many more stars beyond what is formed quiescently, which would make
the expected stellar masses even higher than just from the quiescent mode.

Hence the unseen galaxies must be passive, or forming stars at
some undetectably low rate.  In the currently popular downsizing
scenario~\citep{cow96}, these systems would have formed their stars
earlier in a rapid mode, and have now attenuated their star formation
through some feedback process so as not to be selected in star-forming
samples.  Perhaps by $z\sim 2$, downsizing is already well underway,
and the observed galaxies only represent the small fraction of systems
that have just recently begun forming stars and will soon be quenched.

While qualitatively sensible, the ``unseen passive galaxies" model
is once again quantitatively inconsistent with other observations of
high-$z$ galaxies.  \citet{per07} used a {\it Spitzer}/IRAC-selected
sample  to determine that the star-forming BzK (sBzK) selection technique,
i.e. the one employed by \citet{dad07} to measure $M_*$-SFR at $z\sim 2$, encompasses 88\%
of their galaxies (2691/3044) from $1.6<z<2.5$.  This is remarkably
efficient, significantly more so than e.g. the Lyman break or Distant Red
Galaxy selection techniques in obtaining a stellar mass-limited sample.
Their stellar mass completeness is estimated to be $\sim 75$\% above
$M_*\ga 10^{10.3}M_\odot$ for passively evolving galaxies.  Putting these
together, this means that conservatively at least two-thirds of all
galaxies above this mass threshold are selected as sBzK galaxies, and
hence less than one-third are unseen passive galaxies.  In contrast,
the observed $M_*-$SFR relation at $z\sim 2$ implies $\asf\approx
0.2$, meaning that $\sim 80\%$ of galaxies should be unseen (passive),
or else the duty cycle of such objects should be $\sim 20\%$.
This assumes galaxies began significant activity $\sim 1$~Gyr after the
Big Bang, and that $\asf=0.2$ at $z>2$; the latter is again conservative,
as an extrapolation would indicate that $\asf$ should be even lower
at $z>2$.  Hence it does not appear that sBzK selection misses nearly
enough systems to make the unseen passive galaxies scenario viable
to explain $\asf(z\approx2)$.

Another way to accomodate the low scatter, unity slope, and rapid
evolution in $M_*-$SFR is to propose that galaxies evolve {\it along} the
$M_*-$SFR relation, as proposed by \citet{dad07}.   At $z\sim 2$, $\asf=0.2$
implies that galaxies must be growing very quickly.  Rapid growth means
that for any given galaxy, the amplitude of $M_*-$SFR is essentially
unchanging.  Now since SFR$\propto M_*=\int {\rm SFR}\; dt$, an unchanging
amplitude implies a solution of
\begin{equation}\label{eqn:tau}
{\rm SFR}= C\exp(t/\tau)\; \Longrightarrow \; 
M_*= C\tau\exp(t/\tau)= \tau\; {\rm SFR}.
\end{equation}
In other words, in this scenario, galaxies have exponentially growing star
formation rates and stellar masses!  This growth would presumably continue
until star formation is truncated by some feedback mechanism, probably
when $M_*\ga 10^{11}M_\odot$.  While at first this ``exponentially
growing galaxies" idea may seem incredible, of all the scenarios it
actually comes closest to reconciling all the high-$z$ galaxy data.

One way to test this scenario is to consider the evolution of the number
densities $n$ of galaxies.  The $e$-folding time of stellar mass growth is
given by $\tau=M_*/$SFR.  Assuming (conservatively) that $\asf=0.2$
remains constant at $z\geq 2$, then $\tau \propto t_H-1$~Gyr\;\footnote{
Technically, the fact that $\tau$ changes with time means that the
assumption of a constant $M_*$-SFR amplitude in equation~\ref{eqn:tau}
is invalid; however, a simple evolutionary model accounting for this
yields virtually identical results.}.  A simple calculation shows
that there are e.g. five $e$-folding times (or 2.2 decades of growth)
from $z=3.5\rightarrow 2$.  For a given stellar mass $M_*$, $n(M_*)$
at $z=2$ should then be similar to $n(M_*/10^{2.2})$ at $z=3.5$
(neglecting mergers).  \citet{fon06} measured the number density
of galaxies with $M_*=10^{11}M_\odot$ at $z\sim 2$ to be $7\times
10^{-4}$Mpc$^{-3}$, and those with $M_*=10^{8.8}M_\odot$ at $z=3.5$
have $n=3\times 10^{-3}$Mpc$^{-3}$ (based on an extrapolation of their
STY fits), and stellar mass functions from \citet{els07} yield similar
values.  Hence at face value current data suggests that the stellar mass
function does not evolve quite as rapidly as expected in this scenario.
Still there are substantial uncertainties in the observed stellar mass
function at such low masses and high redshifts, so a factor of a few
discrepancy is not compelling enough to rule out the exponentially
growing galaxies scenario.

Another scenario that has been shown to reproduce $M_*$-SFR evolution is
the staged galaxy formation model of \citet{noe07b}.  Here, a delay is
introduced in the onset of galaxy formation that is inversely related
to the galaxy mass.  Hence large-mass objects form early and quickly,
while small ones start forming later over longer timescales.  This is
similar to the exponentially-growing galaxies scenario in the sense
that the latter implicitly assumes a sort of delay, because galaxies
must accumulate lots of gas in their halo without forming stars in
order to have a sufficient gas reservoir to then consume exponentially.
Hence it may be possible to constrain staged galaxy formation with similar
measurements of the stellar mass function to higher-$z$; this is being
investigated by K. Noeske (priv. comm.).  If either exponentially-growing
or staged galaxy formation is correct, this would indicate that star
formation at early epochs does not scale with gas density as seen
locally~\citep{ken98}, or else that cold gas is somehow prevented from
accumulating in halos as expected from growth of structure.

In summary, the small $\asf$ at $z\sim 2$ implied by the $M_*-$SFR
amplitude is difficult to reconcile with the observed scatter in
$M_*-$SFR, the observed evolution of star-forming galaxies to higher
redshifts, and the observed fraction of passive galaxies.  Taken together,
the latter data all point to the idea that observed star-forming
galaxies represent the majority of galaxies at intermediate masses ($\sim
10^{10}-10^{11}M_\odot$) at $z\sim 2$, and have been quiescently forming
stars over much of a Hubble time, i.e. $\asf\sim 1$.  While designer
models of stellar assembly cannot be excluded, they would be highly
unexpected, and would require substantial revision to our understanding
of how gas accumulates and forms stars in hierarchical structure formation.
The general difficulty in reconciling the data together with the models'
preference for $\asf\sim 1$ suggests an uncomfortable situation that
bears investigation into alternative solutions.

\section{Systematic effects in $M_*$ and SFR determinations}\label{sec:syst}

Determining $M_*$ and SFR from broad-band observations is fraught with
a notoriously large number of systematic uncertainties that could alter
the values of $\asf$ inferred from observations.  Here it is considered
whether such systematics could be responsible for the low value of $\asf$
at high-$z$, particularly focusing on the \citet{dad07} analysis at
$z\sim 2$ since that is where the discrepancies from model expectations
and the range of possible systematics are greatest.  Star formation
rates there are generally estimated by summing rest-UV flux (uncorrected
for extinction) with 24$\mu$ flux.  For objects that are anomalously
high or low in 24$\mu$ to dust-corrected UV SFR estimations, they
use the dust-corrected UV or 24$\mu$ estimate, respectively; however,
only one-third of their sample are anomalous in that sense.  Meanwhile,
stellar masses are estimated from SED fitting: Using multi-band photometry
or spectra, one fits model stellar population templates~\citep[in their
case,][]{bru03}, varying parameters until the best fit is obtained.
In detail, \citet{dad07} used a calibration of BzK color to stellar mass
based on fitting K20 survey spectra~\citep{fon04}.  The key assumptions
in SED fitting are population synthesis models, dust extinction models,
and the form of the star formation history, all of which introduce
systematic uncertainties.

Current uncertainties in obtaining accurate stellar masses from
rest near-IR data revolve around the poorly known contribution of
thermally-pulsating asymptotic giant branch (TP-AGB) stars to the $K$-band
light~\citep{mar06}.  This is found to lower the stellar mass estimates
for high-$z$ galaxies by a factor of two or more~\citep{mar06,bru07}.
However, the sense would be to {\it exacerbate} the problem by lowering
$M_*$ and hence further reducing $\asf$, compared with the \citet{bru03}
models utilized in \citet{fon04}.  Hence while uncertainties in population
synthesis models are not negligible, they seem unlikely to make a
$\ga\times 3$ difference in the proper direction.

It is possible to hide more stellar mass in objects if one postulates a
very old underlying stellar population, in other words by modifying the
assumed SFH form.  \citet{pap01} found that for UV-selected galaxies,
a mass increase of $\sim\times 3-8$ was possible without violating
rest-optical constraints.  However, $K$-selected samples like that
of \citet{dad07} are generally redder with more contribution from
old stellar light, so the amount that one can hide is more limited
(C. Papovich, priv. comm.).  More problematic is the fact that doing so
requires assuming that the bulk of the galaxy's stars formed very early
in the Universe, and hence the current SFR is much lower than at that
early epoch.  It turns out that this is not a self-consistent solution
for reducing $\asf$.  This is because if such a SFH history were true,
one would expect that the current $M_*$ would be significantly {\it more}
than obtained by assuming a constant SFH over a Hubble time.  This would
imply $\asf\gg 1$ for those galaxies.  Hence even hiding $5\times$ as
much stellar mass as currently inferred would be insufficient, because it
would only make $\asf\sim 1$, whereas such a SFH implies $\asf\gg 1$.
In order for there to be a large old stellar population while still
having a high current SFR, it must be that such galaxies formed many
stars early, then sat around not forming stars until a recent flare-up;
however, this scenario suffers the same difficulties as the delayed
galaxy formation model.  Hence one cannot self-consistently alter the
SFHs to hide a large amount of very old stars in these galaxies.

Instead, it may be possible to appeal to some systematic reduction in
the inferred star formation rate.  Why might the true SFR be lower
than estimated?  One possibility is that the extinction corrections
being used for UV-estimated SFRs are too excessive.  Now recall that
for the majority of galaxies, \citet{dad07} estimated the SFR from the
uncorrected UV plus 24$\mu$ flux, so for those cases the extinction
correction is not relevant; it is only relevant for the $\la 25\%$
that are ``mid-IR excess" systems.  For those, there may be room to
alter the extinction law such that the extinction inferred from the UV
spectral slope is overestimated.  \citet{dad07}, like most studies of
actively star forming galaxies, use the \citet{cal00} law.  Using instead
a Milky Way or LMC law would indeed yield less extinction for a given UV
spectral slope, but the typical reduction is less than a factor of two.
So it seems unlikely that extinction alone is causing the low $\asf$
even in the minority mid-IR excess objects, unless the extinction law
were dramatically different than anything seen in the local universe.

Another possibility is that there is an additional contribution to the
rest-UV flux besides star formation.  The obvious candidate are AGN,
which are common in $z\sim 2$ star forming galaxies.  While \citet{pap06}
found that only $\sim 25\%$ of $M_*>10^{11}M_\odot$ galaxies at $z\sim
2$ have X-ray detectable AGN, \citet{dad07b} identified obscured AGN
through their mid-IR excess, and found that they appear in $\sim 50\%$
of galaxies at $M_*>4\times 10^{10}M_\odot$, and dominate the overall AGN
population at $z\sim 2$.  Importantly, they also found \citep[in agreement
with previous studies, e.g.][]{pap06} that smaller mass galaxies have
a lower AGN fraction, as expected theoretically~\citep{hop07}.  Recall
that \citet{dad07} used UV fluxes to estimate SFR in such mid-IR excess
objects, and by comparison with X-ray data they estimated that such an
inferred SFR is unlikely to be wrong by a significant factor.  Moreover,
since AGN are increasingly rare in smaller mass systems, it would be an
odd coincidence if they mimicked the relation $M_*\propto$SFR$^{0.9}$
so tightly.  Like the burst model, this represents a fine-tuning problem,
where AGN have to contribute a very specific amount to the inferred SFR
at each mass scale for no obvious physical reason.  This cannot be ruled
out, but seems a priori unlikely.

It is also possible that AGN could be present in galaxies that are not
mid-IR excess systems.  The strong dichotomy in the X-ray properties
suggest there is not a continuum of AGN, and that AGNs are limited to
mid-IR excess systems~\citep[e.g.][]{dad07b}.  However, there exist
AGN templates where the contibution to rest-UV and rest-8$\mu$ flux
are broadly comparable to that in star-forming galaxies~\citep{dal06}.
These could be mistaken for ``normal" star-forming systems, whereas
in fact the dominant flux is from AGN.  Still, once again, to mimic
the tight $M_*-$SFR relation would require AGN to be contaminating
the SFR estimate at a similar level at all stellar masses.  Moreover,
if they are dominating the UV flux then they should appear as central
point sources, but {\it Hubble}/ACS imaging of sBzK galaxies generally
do not show such bright central sources (M. Dickinson, priv. comm.).
Hence while AGN are certainly present, it seems unlikely that they are
a major contaminant for UV-derived SFRs.

A perhaps even more troublesome issue is the use of rest-8$\mu$
(i.e. 24$\mu$ at $z\sim 2$) as a star formation rate indicator.  This is
quite uncertain, because the dominant flux at those wavelengths is from
polycyclic aromatic hydrocarbon (PAH) emission features, whose nature
is not well understood.  As \citet{smi07} points out, there is more
than a factor of two scatter in star formation rate indicators that
assume a fixed (local) PAH template.  Moreover, owing to PAH features
moving in and out of bands, standard photometric redshift uncertainties
of $\Delta z/(1+z)\sim 0.1$ can result in up to an order of magnitude
error in inferred SFRs.  On the other hand, the tight scatter in the
$z\sim 2$ $M_*-$SFR relation would seem to limit the amount of scatter
owing to PAH feature variations, or else be an indication that UV light
dominates over PAH emission~\citep[though the latter is not what is
seen locally;][]{smi07}.  \citet{cal07} notes that $8\mu$ emission has
a substantial metallicity dependence, but it is in the sense that lower
metallicities have lower emission, so their SFR would be underestimated
by using solar metallicity calibrations.  So while it is conceivable
that some unknown reason causes PAH emission per unit star formation to
be systematically higher in high-$z$ galaxies compared to local ones,
thereby causing the SFR to be overestimated, but there are no obvious
indications that this would mimic a tight $M_*$-SFR relation.

Finally, the observed SFR and $M_*$'s in \citet{dad07} are inferred
assuming a Salpeter IMF.  Assuming instead a \citet{cha03} or
\citet{kro01} IMF would result in shifting {\it both} $M_*$ and SFR down
by $\approx0.15$~dex~\citep{elb07}.  Hence to first order it makes no
difference as long as the $M_*-$SFR slope is around unity (E. Daddi,
priv. comm.).  Fundamentally, this is because at high-$z$ both stellar
mass and star formation rate indicators are driven by light from stars
$\ga 0.5M_\odot$, and favored present-day Galactic IMFs generally agree
on the shape above this mass.  Hence more substantial IMF variations would
be required.

In summary, there are no obvious paths to systematically alter $M_*$ and
SFR determination to make up up the required $\sim\times 3-5$ difference
in $\asf$ at $z\sim 2$.  It is true that many of these systematics
cannot be ruled out entirely, and it may be possible that a combination
of such effects could explain part or even all of the difference.  It is
also possible that locally calibrated relations to estimate $M_*$ and SFR
may be substantially different at high-$z$ for unknown and unanticipated
reasons.  The reader is left to assess the plausibility of such scenarios.
The view taken here is that the difficulty in obtaining a straightforward
solution motivates the consideration of more exotic possibilities.

\section{An evolving IMF?}\label{sec:imf}

One possible way to alter $M_*$-SFR evolution is to invoke an IMF that
is in some direct or indirect way redshift-dependent.  The key point
is that most direct measures of the star formation rate actually trace
high-mass star formation, while stellar mass is dominated by lower-mass
stars.  Hence by modifying the ratio of high to low mass stars formed,
i.e. the shape of the IMF, it is possible to alter the $M_*-$SFR relation.
The possibility of a varying IMF has been broached many times in various
contexts~\citep[e.g.][]{lar98,fer02,lac07,far07}, though typically as a
last resort scenario.  It is highly speculative as there is no clear-cut
evidence at present that supports a time- or space-varying IMF, but it
is worth investigating in light of growing observational and theoretical
arguments in its favor.

As \citet{kro01} points out, a universal IMF is not to be expected
theoretically, though no variations have been unequivocally detected in
local studies of star-forming regions.  The average Galactic IMF appears
to be well-represented by several power laws, steepest at $>1 M_\odot$,
then significantly shallower at $<0.5 M_\odot$, with a turnover at
$<0.1 M_\odot$~\citep[see review by][]{kro07}.  Forms that follow this
include the currently-favored \citet{kro01} and \citet{cha03} IMFs.
But deviations {\it are} seen, at least at face value.  The nearby
starburst galaxy M82 appears to have an IMF deficient in low-mass
stars~\citep{rie80,rie93}.  A similarly bottom-light IMF is inferred in
the highly active Arches cluster near the Galactic center, suggesting that
vigorous star formation makes the IMF top-heavy~\citep[e.g.][]{fig05}.
Also, the young LMC cluster R136 shows a flattening in the IMF below
$\sim 2M_\odot$~\citep{sir00}.  There are hints that presently active
star clusters form more low-mass stars compared to the disk-averaged 5
Gyr-old IMF~\citep{kro07}, suggesting that the IMF was weighted towards
heavier stars at lower metallicities and/or earlier times.  Even if
the IMF shape is universal, larger starforming regions may produce more
high-mass stars simply because they have enough material to aggregate
into larger stars~\citep{wei06}.

Hence there are hints of IMF variations, albeit controversial, and
interestingly they consistently go in the direction of having a higher
ratio of massive to low-mass stars in conditions similar to those in
high-$z$ galaxies.  Compared to present-day star-forming systems, galaxies
at high-$z$ have far higher star formation surface densities than the
Galactic disk~\citep{erb06c}, have higher gas content and hence presumably
more massive starforming clouds~\citep{erb06b}, and have somewhat lower
metallicities~\citep{sav05,erb06}.  \citet{lar98} and \citet{far07}
list other pieces of circumstantial evidence that favor early top-heavy
IMFs.  So while there are no direct observational evidences or strong
theoretical arguments for it, it is not inconceivable that the IMF in
high-$z$ galaxies may be more top-heavy or bottom-light\footnote{The
distinction between top-heavy and bottom-light IMFs is mostly semantic.
Here the convention is used that top-heavy means a high-mass slope that
is less steep than the local Salpeter value, while bottom-light means
that the high-mass slope remains similar but low-mass stars are
suppressed.} than the present-day Galactic IMF.

Such an IMF would have three main effects on the $M_*-$SFR relation:
It would increase the output of UV flux per unit stellar mass formed; it
would cause more stellar mass loss from stellar evolution; and the mass
recycling would provide a larger gas reservoir with which to form stars.
All three effects go towards reconciling theoretical expectations with
observations of the $M_*$-SFR relation.  The SFR inferred by assuming a
standard IMF would be an overestimate, since many fewer low-mass stars
are formed per high-mass star.  Increased recycling losses would lower
the theoretical expectations for the amount of stellar mass remaining
in these galaxies, bringing them more into line with observed stellar
masses.  Recycled gas would increase the available reservoir for new
stars, effectively delaying star formation in galaxies and pushing
models towards the ``delayed galaxy formation" scenario.  Hence even
modest evolution in the IMF could in principle reconcile the observed
and theoretically-expected $\asf$ evolution.

\section{A Simple IMF Evolution Model}\label{sec:imfmodel}

How is the IMF expected to evolve with time?  This is difficult to
predict since no ab initio theory for the IMF exists today.  But a simple
conjecture by \citet{lar98,lar05} may have some empirical value.  He notes
that there is a mass scale of $\sim 0.5 M_\odot$ below which the Galactic
IMF becomes flatter~\citep{kro93,kro01,cha03}, and at which point the
mass contribution per logarithmic bin ($M dN/d\log{M}$) is maximized.
Larson uses simple thermal physics arguments to show that this {\it
characteristic IMF mass scale} ($\hat{M}_{\rm IMF}$) is related to the
minimum temperature $T_{\rm min}$ of molecular clouds, measured today
to be around 8~K.  In that case, if the temperature in the interstellar
media (ISM) of high-$z$ galaxies is higher, this characteristic mass
will shift upwards, resulting in an IMF that is more bottom-light.

To generate a crude IMF evolution model, let us assume that the
$\hat{M}_{\rm IMF}\propto (1+z)^{\gamma}$, where $\gamma$ is some unknown
parameter whose value is set by the various effects above.  \citet{lar85}
predicted analytically that $\hat{M}_{\rm IMF}\propto T_{\rm min}^{3.35}$,
while recent numerical experiments by \citet{jap05} found a less steep
scaling of $T_{\rm min}^{1.7}$.  So for example an evolution that scaled
purely with CMB temperature would yield $\gamma=1.7$.  However, other
factors may cause an increased ISM temperature, e.g., elevated disk SFR
rates causing more supernova heat input; lower cooling rates owing to
lower metallicities; and/or photoionization from a stronger metagalactic
UV flux~\citep[e.g.][]{sco02}.

The approach adopted here is to treat $\gamma$ as a free parameter, to
be constrained using the observed evolution of $\asf$.  The assumption is
that $\asf(z=0)$ reflects the present-day Galactic IMF, while $\asf(z>0)$
is inferred to be lower purely because of the assumption of a non-evolving
IMF, whereas in reality $\asf(z)$ (at any given mass) is constant as
expected in models.

An ``evolving Kroupa" IMF is defined in which
\begin{equation} \label{eqn:zkroupa}
\hat{M}_{\rm IMF}= 0.5 (1+z)^{\gamma} M_\odot.
\end{equation}
Above $\hat{M}_{\rm IMF}$, the IMF has a form
$dN/d\log{M}\propto M^{-1.3}$, while below it scales as $M^{-0.3}$.
Note that the current Galactic IMFs also have a turn-down at masses 
below $0.1 M_\odot$;
the present analysis provides no constraints in this regime, as the
stellar mass formed there is a small fraction of the total.
The exact form of the IMF is not critical here; \citet{kro01} is chosen for
its convenient explicit parameterization in terms of $\hat{M}_{\rm IMF}$,
but the analysis below could equally well have
been done with the similar Chabrier IMF.

The goal of the evolving IMF is to increase the ratio of the star
formation rate of {\it high-mass} stars (SFR$_{\rm hiM}$) to total
stellar mass formed, such that the inferred $\asf(z)$ is non-evolving.
Note that different probes such as UV luminosity, H$\alpha$ emission,
and mid-IR emission trace slightly different regimes of high-mass star
formation, but so long as the high-mass IMF slope remains fixed, the
relative ratios of such emission should be constant.  So for example,
in order to produce no evolution in $\asf$ from $z\sim 0\rightarrow 2$
at $10^{10}M_\odot$, the factors by which $f\equiv$SFR$_{\rm hiM}/M_*$
must be raised are 1.9 at $z=1$ and 3.3 at $z=2$, relative to a standard
Kroupa IMF (cf. Figure~\ref{fig:tact}).  Hence $\gamma$ must be chosen
to match these factors.

What value of $\gamma$ would produce such evolution?  To determine this,
the PEGASE.2~\citep{fio97} population synthesis model is employed, taking
advantage of its feature allowing user-definable IMFs.  PEGASE outputs
a variety of quantities as a function of time from the onset of star
formation, given an input star formation history.  The assumed SFH is
taken to be exponentially-declining with a 10~Gyr decay time and starting
1~Gyr after the Big Bang, in order to approximate a typical model SFH.
An example of such a SFH is shown as the magenta line in the lower right
panel Figure~\ref{fig:sfrmstar}, normalized to a star formation rate
of 2~$M_\odot$yr$^{-1}$ today; as is evident, it is quite similar to
typical SFHs from the simulations, at least for moderate stellar masses.
A constant SFH was also tried, and the results were not significantly
different.  Besides the input SFH, all other parameters are taken as
the PEGASE default values.  No dust extinction is assumed.

The PEGASE output quantities specifically used here are the stellar mass
$M_*$ formed, and the un-extincted 1650\AA\ luminosity ($L_{1650}$).
The latter is used as a proxy for SFR$_{\rm hiM}$.  Using these, it is
possible to compute $f$ for any given assumed IMF at any given time.
Note that PEGASE accounts for both stellar mass accumulation and mass
loss owing to stellar evolution, but does not account for the impact of
recycled gas.  By varying $\hat{M}_{\rm IMF}$, it is possible to estimate
$f$ for any assumed $\gamma$, and thereby determine what value of $\gamma$
would yield no $\asf$ evolution.

In practice, determining $f\equiv$SFR$_{\rm hiM}/M_*$ at any epoch requires
integrating the mass growth up to that epoch for an evolving IMF.  To do
so, first PEGASE models are computed for a wide range of $\hat{M}_{\rm
IMF}$ values.  Then the following procedure is used:\\
(1) A value of $\gamma$ is selected; \\
(2) the following integral is computed:
\begin{equation} \label{eqn:mchar}
M_*(t_0) = \int_0^{t_0} \frac{\Delta M}{\Delta t}(\hat{M}_{\rm IMF},t) dt,
\end{equation}
where $\Delta M/\Delta t$ is the stellar mass growth rate at time $t$
taken from PEGASE models interpolated to the appropriate $\hat{M}_{\rm
IMF}$ as given in equation~\ref{eqn:zkroupa}; \\
(3) $L_{1650}$ is taken at time $t$ from PEGASE, 
and $f_{\rm ev}(t_0)=L_{1650}(t_0)/M_*(t_0)$ is computed (the subscript
``ev" stands for ``evolving"); \\
(4) Similarly, $f_{\rm st}(t_0)$ is computed
for a standard Kroupa IMF, where $\hat{M}_{\rm IMF}=0.5 M_\odot$;\\
(5) The ratio $f_{\rm ev}(t_0)/f_{\rm st}(t_0)$ is computed, 
and compared to the factors needed to produce no $\asf$ evolution.

\begin{figure}
\vskip -0.4in
\setlength{\epsfxsize}{0.65\textwidth}
\centerline{\epsfbox{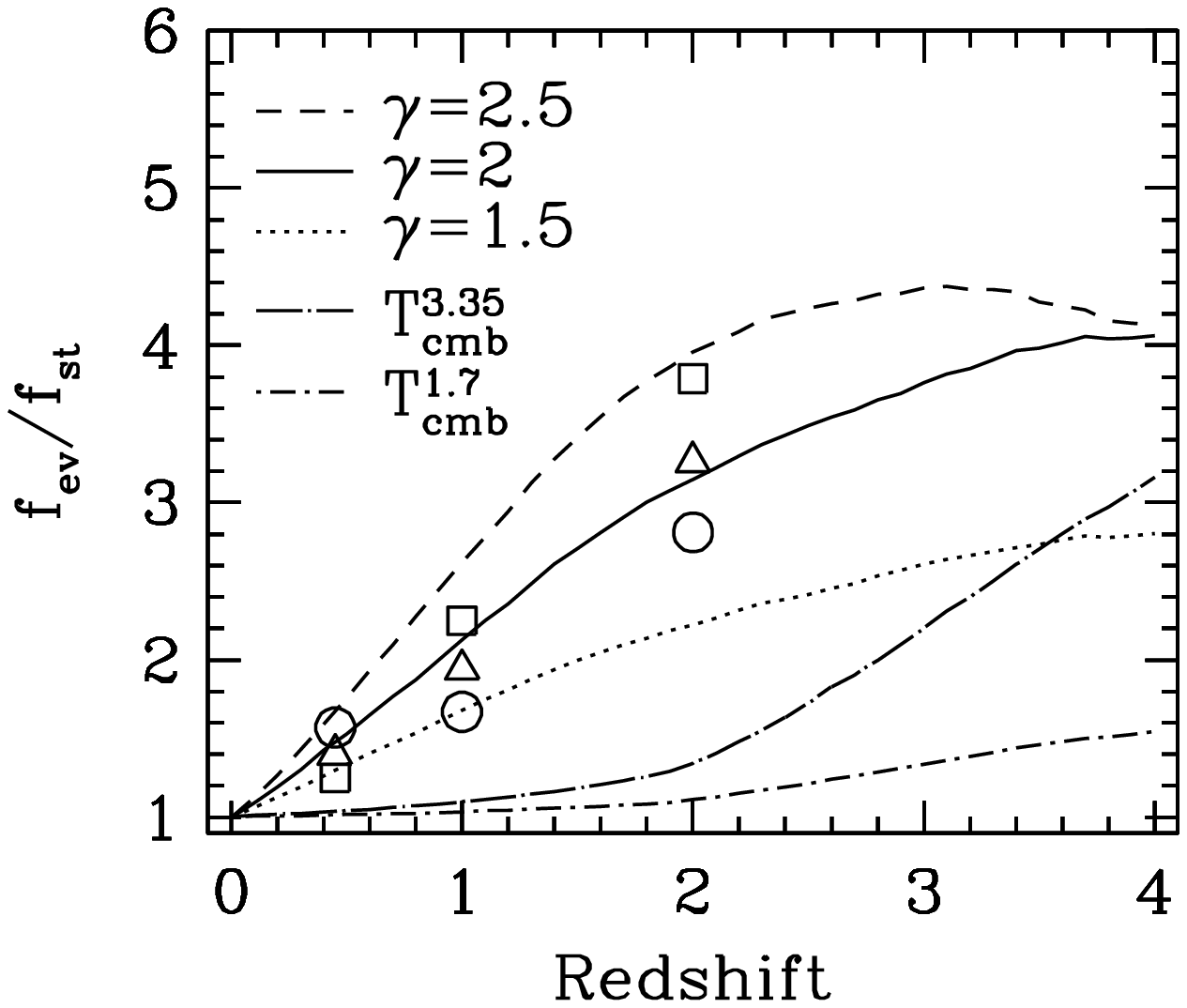}}
\vskip -3.2in
\caption{Evolution of the ratio of high-mass star formation rate to
stellar mass, for the evolving Kroupa IMF (equation~\ref{eqn:zkroupa})
relative to a standard Kroupa IMF, normalized to unity at $z=0$.  The
curves are computed from PEGASE.2 models for different values of $\gamma$,
as described in \S\ref{sec:imfmodel}.
The data points are from fits to the observed $M_*-$SFR relation at
$z=0.45,1,2$ (described and shown in Figure~\ref{fig:tact}) at stellar
masses of $10^{10.5}M_\odot$ (squares), $10^{10}M_\odot$ (triangles), and
$10^{9.5}M_\odot$ (circles).  
The required evolution is reasonably well fit by $\gamma=2$, 
i.e. $\hat{M}_{\rm IMF}= 0.5 (1+z)^{2} M_\odot$.  Long and short
dashed-dot models show results for the \citet{lar98} scenario where
the minimum temperature of molecular clouds is set by the CMB
temperature, and $\hat{M}_{\rm IMF}\propto T_{\rm min}^\beta$, where
$\beta=3.35$~\citep{lar85} and $\beta=1.7$~\citep{jap05}, respectively.
Because the CMB temperature does not exceed the the local minimum
temperature in molecular clouds of 8~K until $z\ga 2$, this scenario
does not yield substantial evolution at $z\la 2$.
}
\label{fig:zratio}
\end{figure}

The results of this procedure are shown in Figure~\ref{fig:zratio},  
namely $f_{\rm ev}/f_{\rm st}$ as
a function of redshift for several values of $\gamma$. 
The values by which the $M_*-$SFR
relation needs to be adjusted to produce no $\asf$ evolution at
$z=0.45,1,2$ are indicated by the symbols: Squares are for
$M_*=10^{9.5}M_\odot$, triangles for $10^{10}M_\odot$, and circles 
for $10^{10.5}M_\odot$.  These are computed from published fits to $M_*-$SFR 
\citep{noe07a,elb07,dad07}.  Because the $M_*-$SFR relation
becomes shallower with time, the required adjustment factors are 
generally smaller at lower masses.

Figure~\ref{fig:zratio} shows that $\gamma=2$ provides a reasonable
fit to the required IMF evolution for typical star-forming galaxies at
$M_*=10^{10}M_\odot$.  The implied values of $\hat{M}_{\rm IMF}$ are $2
M_\odot$ and $4.5 M_\odot$ at $z=1,2$, respectively.  Note that there
is no a priori reason why a single value of $\gamma$ should fit at all
redshifts considered, so the good fit supports the form of $\hat{M}_{\rm
IMF}$ evolution assumed in equation~\ref{eqn:zkroupa}.

As a check, an independent method for determining $\hat{M}_{\rm IMF}$
evolution was explored that is less comprehensive, but conceptually more
straightforward.  Here, the high-mass star formation rate is assumed to be
unchanged, while observations of stellar mass are assumed to reflect stars
near the main sequence turnoff mass ($M_{\rm turnoff}$), since red optical
light is dominated by giant stars.  For an evolving Kroupa IMF, to obtain
a reduction of a factor of $x$ in the amount of stars produced at $M_{\rm
turnoff}$, it is easy to show that $\log{\hat{M}_{\rm IMF}}=\log{M_{\rm
turnoff}}+\log{x}$.  $M_{\rm turnoff}$ may be estimated by noting that
stellar lifetimes scale as $M^{-3}$, and that the Sun has a lifetime of
10~Gyr; this yields $M_{\rm turnoff}^{3}=10/(t_H-1)$, ($t_H$ in Gyr),
which is $1.24M_\odot$ and $1.58M_\odot$ at $z=1,2$, respectively.
Inserting values of $x=1.9,3.3$ at $z=1,2$ yields $\hat{M}_{\rm IMF}=2.4$
and $5.2 M_\odot$.  These values are quite similar to those obtained from
PEGASE modeling, showing that the results are not critically dependent
on details of PEGASE.

The same procedure can be applied to the scenario proposed by
\citet{lar98}, where the CMB temperature is solely responsible for
setting a floor to the ISM temperature.  In that case, $T_{\rm min}={\rm
MAX}[T_{\rm CMB},8 {\rm K}]$, and $\hat{M}_{\rm IMF}\propto T_{\rm
min}^{\beta}$.  \citet{lar85} predicted $\beta=3.35$, while \citet{jap05}
found $\beta=1.7$.  The results of this scenario for these two values of
$\beta$ are shown as the long and short dot-dashed lines, respectively, in
Figure~\ref{fig:zratio}.  As is evident, this IMF evolution is not nearly
sufficient to reconcile the observed and predicted $\asf$ evolution out to
$z\sim 2$, mainly because $T_{\rm CMB}$ only exceeds 8~K at $z>1.93$, and
hence $\hat{M}_{\rm IMF}$ does not change from $z\sim 0-2$.

Note that since CMB heating of the ISM is subdominant at all redshifts,
the assumed form of IMF evolution (eq.~\ref{eqn:zkroupa}, scaling
as $T_{\rm CMB}^\gamma$) is not physically well motivated.  While the
evolution out to $z\sim 2$ is well-fit by such a form, it could be that
the form changes at higher $z$, or that it should be parameterized as
some other function of $z$.  For instance, if it is the vigorousness of
star formation activity that determines $\hat{M}_{\rm IMF}$, perhaps IMF
evolution actually reverses at very high redshifts.  For lack of better
constraints, the form for IMF evolution in equation~\ref{eqn:zkroupa}
will be assumed at all $z$, but this should be taken as an illustrative
example rather than a well-motivated model.

This analysis also makes predictions for $\asf$ at $z>2$ that would
be inferred assuming a standard (non-evolving) IMF.  For instance,
Figure~\ref{fig:zratio} implies that at $10^{10}M_\odot$, $\asf(z=3)=0.13$
and $\asf(z=4)=0.12$.  Notably, the evolution slows significantly at
$z\ga 3$, and actually reverses at $z\ga 4$, though the simple SFH
assumed in the PEGASE modeling may be insufficiently realistic to yield
valid results at $z\ga 4$.  Any such extrapolation of this IMF evolution
should be made with caution, as the $M_*$-SFR relation only constrains
it out to $z\sim 2$.  However, in the next section this will be compared
to observations that suggest that such IMF evolution may be reasonable
out to $z\sim 3+$.

\begin{figure}
\vskip -0.4in
\setlength{\epsfxsize}{0.65\textwidth}
\centerline{\epsfbox{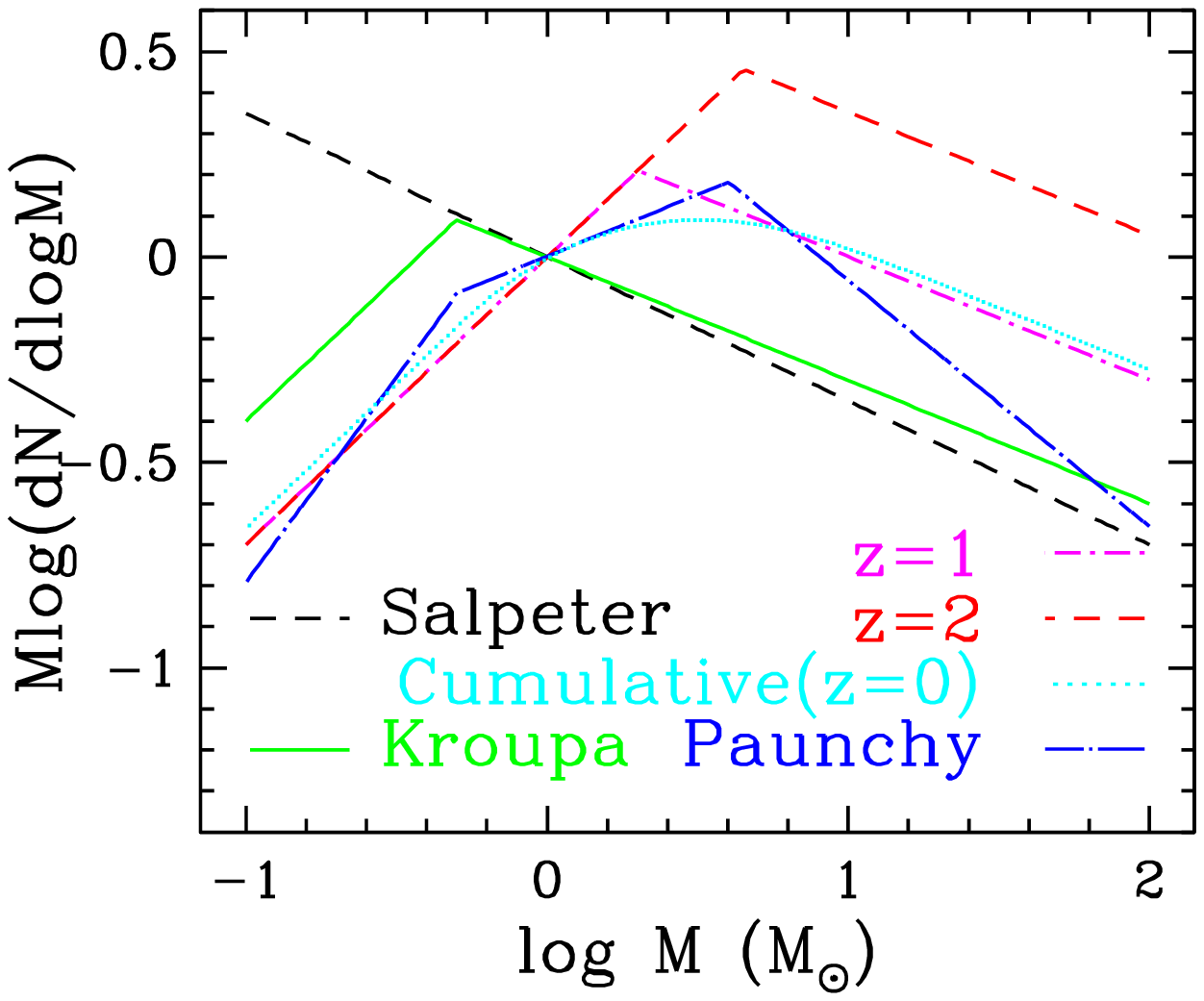}}
\vskip -3.2in
\caption{Initial mass functions, multiplied by $M$ to show the mass
formed per unit logarithmic mass bin, and arbitrarily normalized at
1$M_\odot$.  Salpeter (dashed) and \citet{kro01} (green solid) IMFs are shown;
Chabrier is similar to Kroupa.  Evolving Kroupa IMFs are shown at
$z=1$ (magenta got-dashed) and $z=2$ (red dashed), with characteristic
masses $\hat{M}_{\rm IMF}=2 M_\odot$ and $4.5 M_\odot$, respectively,
as required to remove evolution in $\asf$ (see \S\ref{sec:imfmodel}).  The cumulative $z=0$
IMF (cyan dotted) shows the summation of an evolving Kroupa IMF 
with $\hat{M}_{\rm IMF}=0.5(1+z)^{2}$ over the
history of the Universe, assuming an evolving IMF cosmic star formation
history as shown in Figure~\ref{fig:madau}.  Also shown is the ``paunchy"
IMF (blue) favored by \citet{far07} for reconciling the cosmic star
formation history, the present-day $K$-band luminosity density, and
extragalactic background light constraints.  The cumulative $z=0$
evolving IMF is similar to the paunchy IMF, showing that it
would go well towards reconciling fossil light data with the integrated
cosmic SFH (see \S\ref{sec:mchar}).
}
\label{fig:mchar}
\end{figure}

Figure~\ref{fig:mchar} illustrates the evolving IMF, in terms of stellar
mass formed per unit logarithmic mass bin.  A Salpeter IMF is shown
for reference, and evolving Kroupa IMFs are shown at $z=0,1,2$ with
characteristic masses $\hat{M}_{\rm IMF}=0.5 M_\odot$, $2 M_\odot$,
and $4.5 M_\odot$, respectively.  As expected, the evolution results in
more high mass stars compared to low mass ones at higher redshifts.
While such a dramatic increase in $\hat{M}_{\rm IMF}$ may seem surprising,
recall that studies of local highly active star-forming regions suggest a
characteristic mass scale of a few $M_\odot$~\citep{rie93,sir00,fig05}.
Note that an IMF with a steeper high-mass slope~\citep[e.g.][]{mil79,kro93}
would result in even stronger IMF evolution being required than in the
\citet{kro01} case.

In summary, an evolving Kroupa IMF whose characteristic mass increases
with redshift as $\hat{M}_{\rm IMF}=0.5 (1+z)^{2} M_\odot$ is able to alter
the observed $\asf$ evolution from $z\sim 2\rightarrow 0$ into one
with approximately no evolution, as predicted by models or inferred
from complementary observations.  The precise IMF shape is not well
constrained; it is certainly possible that an IMF with different behavior,
for instance an evolving high-mass slope~\citep[e.g.][]{bal03}, could
produce the same $\asf$ evolution.  The only requirements are that it
produces the correct ratio of high-mass stars relative to low-mass stars
as a function of redshift, and that \citep[unlike e.g. the merger-induced
top-heavy IMF suggested by][]{lac07} it applies to the bulk of star
forming galaxies at any epoch.

\section{Implications of an Evolving IMF}\label{sec:mchar}

What are the observational implications of such an evolving IMF
at early epochs?  Surprisingly few, as it turns out, so long as
the massive end of the IMF remains unaffected.  This is because, as
mentioned before, virtually all measures of high-$z$ star formation, be
they UV, IR or radio, trace light predominantly from high-mass stars.
Feedback energetics and metallicities likewise reflect predominantly
massive star output.  Hence so long as an IMF preserves the same
high mass star formation rate, it is expected to broadly preserve
the successes of understanding the relationship between the rest-UV
properties of galaxies~\citep[e.g.][]{dav06,bou07}, cosmic metal
pollution~\citep[e.g.][]{opp06,dav07}, feedback strength, and Type~II
supernova rates.

The main difference caused by such IMF evolution is in stellar mass
accumulation rates.  High-$z$ measures of stellar mass are therefore
critical for testing this type of scenario.  \citet{dro05} determined the
stellar mass function evolution from a K-band survey out to $z\sim 5$,
but at even moderately high redshifts the rest-frame light is actually
quite blue, so their stellar mass estimates implicitly involves a
significant IMF assumption.  The high-$z$ stellar mass--metallicity
relation~\citep{erb06} has the potential to test the IMF; if the
evolving IMF is correct, the good agreement found versus recent
models~\citep{fin07b,bro07,kob07} would perhaps not stand, and in
particular many more metals would have to be driven out of galaxies of
a given $M_*$ via outflows~\citep[i.e. the ``missing metals" problem
would get worse;][]{dav07}.  Type~Ia supernova rates at high-$z$ would
be lowered; the apparent turn-down in Type~Ia rates at $z\ga 1$ argued by
\citet{hop06} as being indicative of a long Type~Ia delay time of 3~Gyr,
may instead be accomodated via a more canonical short delay time coupled
with an evolving IMF.  These are examples of the types of observations
that could potentially constrain IMF evolution.

\begin{figure}
\vskip -0.3in
\setlength{\epsfxsize}{0.65\textwidth}
\centerline{\epsfbox{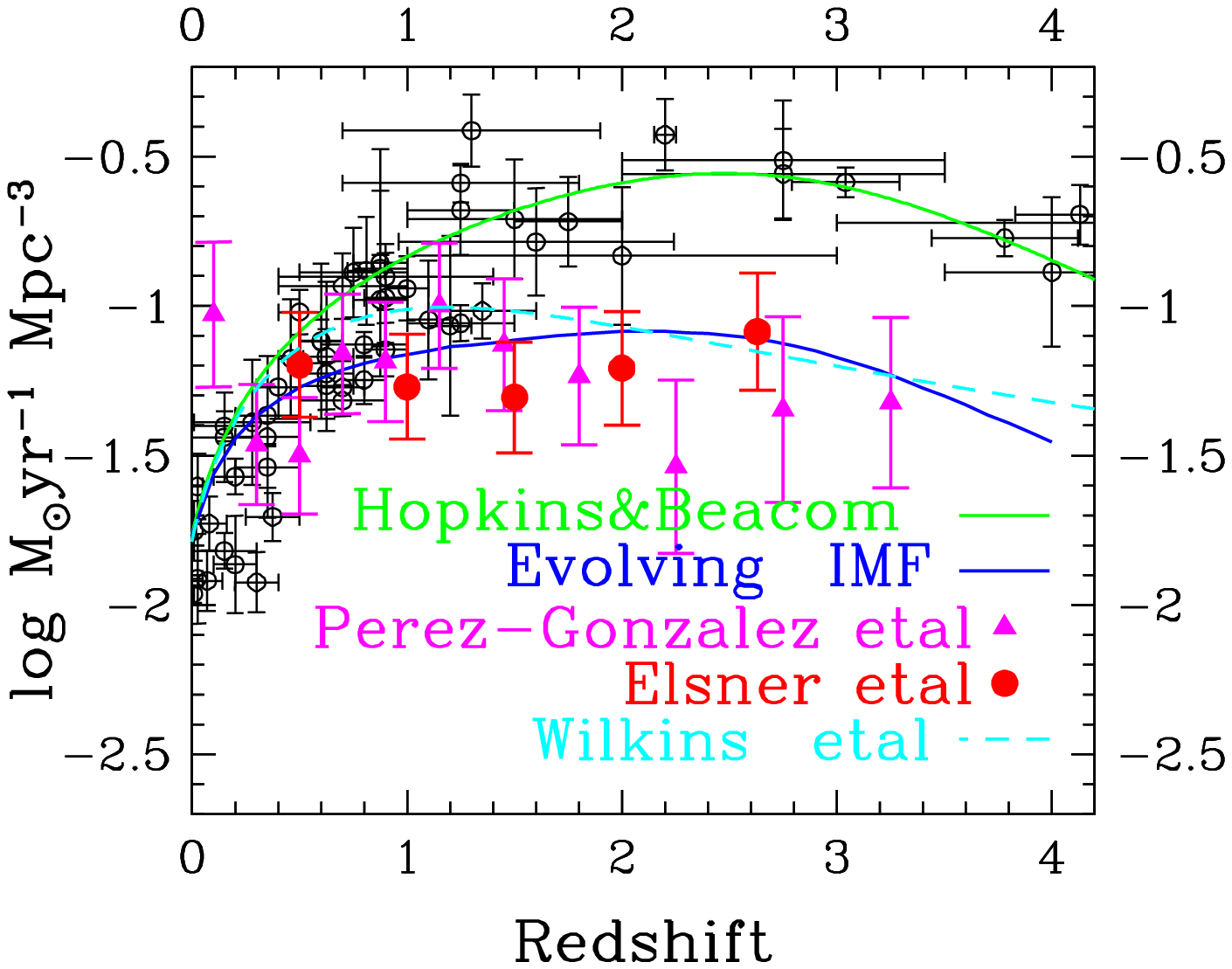}}
\vskip -3.2in
\caption{Cosmic star formation history.  Data points are from
\citet{hop04}; best-fit relation to an updated compilation 
from \citet{hop06} is shown as the green line.  
A schematic representation of the cosmic star
formation history assuming an evolving IMF with $\gamma=2$ is shown as
the blue line, which is simply the \citet{hop06} fit divided
by the ratio of high-mass star formation to stellar mass growth for
an evolving IMF versus a constant one, namely the solid ($\gamma=2$)
line in Figure~\ref{fig:zratio}.  Magenta and red points show the result of
differentiating the cosmic stellar mass density measurements as a
function redshift by~\citet{per07} and \citet{els07}, respectively.  
Cyan dashed line shows the fit to cosmic
star formation rate inferred from a compilation of cosmic stellar mass
densities by \citet{wil07}.  
For ease of comparison, all data here are
shown here assuming a Salpeter IMF.  The evolving IMF case is in good 
agreement with direct measures of stellar mass growth.
}
\label{fig:madau}
\end{figure}

The star formation rates inferred for high-redshift galaxies by assuming
present-day IMFs would have to be revised downwards for an evolving IMF.
A quantitative estimate of the reduction in star formation rates from
the standard to evolving IMF case is provided by the ratio $f_{\rm
ev}/f_{\rm st}$ as a function of redshift from Figure~\ref{fig:zratio}.
Specifically, the ``true" cosmic star formation rate (assuming the
evolving IMF is correct) is the one currently inferred assuming a
non-evolving IMF, divided by the curve for a particular value of $\gamma$,
say $\gamma=2$.

Figure~\ref{fig:madau} shows the cosmic star formation history with data
points compiled by \citet{hop04}, and the fit by \citet{hop06}.  It also
schematically illustrates the effect of an evolving IMF by dividing the
\citet{hop06} fit by the $f_{\rm ev}/f_{\rm st}$ curve for $\gamma=2$.
Note that \citet{hop06} used a Salpeter IMF to infer star formation rates,
so these correction factors (derived for a Kroupa IMF) are not exact,
but should be reasonably close.  Figure~\ref{fig:madau} shows that
the cosmic SFR peak shifts to later epochs in the evolving IMF case.
The total stellar mass formed in the Universe is reduced by 55\% by $z=0$
and 75\% by $z=2$, compared to a standard IMF.

Interestingly, some current observations seem to {\it prefer} an IMF at
early epochs that is more top-heavy or bottom-light, and perhaps a later
epoch of cosmic star formation than generally believed.  \citet{hop06}
noted that the integral of the cosmic star formation history exceeds the
stellar mass currently detected by a factor of two, assuming a Salpeter
IMF.  \citet{bor06} also noted such tension, but suggested that switching
from Salpeter to a Chabrier or Kroupa IMF might go significantly towards
reconciling the difference; little evidence exists for IMF evolution
to $z\la 1$~\citep{bel07}.  But to higher redshifts, observations by
\citet{rud06} of stellar mass density evolution to $z\sim 3$ suggest that
mass buildup is peaked towards significantly later epochs than in current
models that broadly match the observed cosmic star formation history.
\citet{bal03}, updating an analysis done by \citet{mad98}, found that the
cosmic SFH and present-day cosmic luminosity density are best reconciled
using an IMF slope that is slightly more top-heavy than Salpeter ($-1.15$
vs. $-1.35$), and significantly more top-heavy than e.g. \citet{kro93}.
\citet{hea04} suggests from an archaeological analysis of SDSS spectra
that the cosmic star formation rate peaked at $z\sim 0.6$ rather than
the more canonical $z\sim 2$.  \citet{van07} find that it is easier
to understand early-type fundamental plane evolution at $z\la 1$
with a slightly top-heavy IMF ~\citep[as also argued by][]{ren05}.
Intracluster medium metallicities indicate that stars formed in clusters
at early epochs either had higher yields than today, or else formed with
a top-heavy IMF~\citep{por04}.  \citet{luc05} and \citet{tum07} suggest
that the abundance of carbon-enhanced metal-poor stars is indicative
of a more top-heavy IMF at early times in our Galaxy.  \citet{cha07}
finds that an IMF slope more top-heavy than Salpeter is required to
produce enough early photons to reionize the Universe.  While it should
be pointed out that a large number of observations are consistent with
a universal IMF, whenever a discrepancy exists, it is invariably in the
sense of favoring an IMF with more massive stars at early times.

\citet{far07} did perhaps the most careful work on trying to reconcile
fossil light measures with redshift-integrated measures of stellar mass.
Extending the analyses of \citet{bal03} and \citet{hop06} by including
extragalactic background light (EBL) constraints, they found that {\it no}
standard IMF (Kroupa, Chabrier, Salpeter, etc.), is able to reconcile
at the $\sim 3\sigma$ level the cosmic star formation history, the
present-day $K$-band luminosity density, and the EBL.  They examined how
the IMF might be altered to make these observations self-consistent, and
found that what was required was an IMF that had an excess at intermediate
masses ($\sim {\rm few}\; M_\odot$) over standard IMFs.  This IMF need
not be universal, but instead could be a cumulative IMF of all stars
that have formed in the Universe, since it was obtained from fossil light
considerations.  An example of a concordant IMF is their ``paunchy" IMF,
shown as the blue dot-dashed line in Figure~\ref{fig:mchar}.

Using the evolving Kroupa IMF, it is possible to compute a cumulative
IMF of all star formation in the Universe.  This is done by integrating
the evolving IMF over cosmic time weighted by the cosmic star formation
history, which is taken to be the evolving IMF one (blue line)
from Figure~\ref{fig:madau}.  The result is shown as the dotted cyan
line in Figure~\ref{fig:mchar}.  It is similar to the paunchy IMF
of \citet{far07}, showing elevated formation rates of intermediate
mass stars, though it is slightly more top-heavy.  This comparison is
preliminary because the EBL and current $K$-band light trace stellar
output differently than the cosmic star formation history.  Still, it is
likely that this evolving IMF will go significantly towards reconciling
the various observations considered in \citet{far07}.

\citet{van07b} recently argued for a more top-heavy IMF at early times
from the evolution of colors and mass-to-light ratios of early-type
galaxies in clusters from $z\sim 0.8\rightarrow 0$.  Parameterizing the
evolution in terms of the characteristic mass of a Chabrier IMF, he
favored a characteristic mass that evolves by $\times 20$ from $z\sim
4$ until today; this is consistent with $\hat{M}_{\rm IMF}\propto
(1+z)^2$ as proposed here.  Note that his analysis only applies to
stars formed in early-type cluster galaxies, so is not necessarily
applicable to all galaxies.  Making the assumption that it is widely
applicable, \citet{van07b} determined a cosmic stellar mass growth
rate that is broadly consistent with the evolving IMF case shown in
Figure~\ref{fig:madau} (though in detail it is more peaked towards lower
$z$; see his Fig.~13).  While many uncertainties are present in the his
analysis, it is reassuring that similar IMF evolution is inferred
from a completely independent line of argument.

The ultimate test of an evolving IMF is to directly measure the
stellar mass buildup in the Universe.  Recently, \citet{per07} and
\citet{els07} did just that using deep {\it Spitzer}/IRAC observations
that probe near-IR light all the way out to $z\sim 3+$.  Their data of
cosmic stellar mass mass growth rates as a function of redshift are
shown as the magenta and red data points in Figure~\ref{fig:madau},
respectively (assuming a Salpeter IMF).  These data points were obtained
by differentiating the stellar mass densities as a function of redshift
published in those papers.  Although those two samples are independent,
both show that the stellar mass growth rate is significantly different
than that inferred from the observed cosmic star formation history,
particularly at $z\ga 2$.  \citet{per07} notes that this may favor an
IMF weighted more towards massive stars at high redshifts.  They mention
that utilizing a Chabrier or Kroupa IMF will lessen the discrepancy,
but such a reduction is essentially by a fixed factor at all redshifts
(modulo minor stellar evolution differences), so even if the normalization
is adjusted to agree with the cosmic SFH at $z=0$, it would still yield
significantly lower SFRs at $z\ga 2$.

A major uncertainty in this comparison is that \citet{per07}, in
order to integrate up the total stellar mass, assumed a \citet{sch76}
faint-end slope that is fixed at $\alpha=-1.2$, even though they cannot
directly constrain $\alpha$ at $z\ga 2$.  They justify this assumption
by noting that there is no evidence for a change in $\alpha$ from $z\sim
0\rightarrow 2$.  But if $\alpha$ becomes progressively steeper with
redshift, as suggested by \citet{fon06}, it could also reconcile these
observations.  Note that it would have to steepen quickly, e.g. to
$\alpha\approx -1.8$ at $z\sim 3$ to reconcile the $\sim\times 3$
difference, significantly faster than $d\alpha/dz=-0.082$ observed by
\citet{fon06} out to $z\sim 4$.  \citet{els07} also found that $\alpha$
does increase with redshifts slightly in one form of their analysis,
though again not by enough to explain the discrepancy.  Still, the
difficulty of constraining the faint-end slope at high-$z$, along with
inherent uncertainties in SED fitting, mean that these observations
cannot be considered definitive proof of a varying IMF.  But at least
at face value the agreement with the evolving IMF case is quite good.

\citet{wil07} independently determined the cosmic stellar mass
growth rate by compiling observations of stellar mass densities from
the literature for $z\sim 0-4$~(note that they do not include the
Perez-Gonzalez et al. or Elsner et al. data).  Carefully accounting
for various different systematics, they fit a \citet{col01} form to the
implied cosmic star formation history (including effects of stellar mass
loss), and determine $\dot{\rho_*}=(0.014+0.11z)h[1+(z/1.4)]^{2.2}$,
in $M_\odot$yr$^{-1}$Mpc$^{-3}$.  This relation is shown as the
cyan dashed line in Figure~\ref{fig:madau}; it has been adjusted by
$+0.22$~dex to change from the IMF they assumed (similar to Chabrier)
to a Salpeter IMF.  It agrees well with other observations, and lies
significantly below the star formation rate inferred from high-mass SFR
indicators at high-$z$.  Wilkins et al. suggest that an evolving IMF may
be the cause, and propose their own form of IMF evolution to explain it.
As seen in Figure~\ref{fig:madau}, the agreement with the evolving IMF
proposed here is quite good.

In short, an evolving Kroupa IMF does not obviously introduce any large
inconsistencies with high-$z$ galaxy data, and moreover may be preferred
based on low-redshift fossil light considerations and observed cosmic
stellar mass growth rates.  Hence the amplitude evolution of the $M_*-$SFR
relation may be yet another example of an observation of stellar mass
buildup that suggests that the IMF is weighted towards more massive
stars at higher redshifts.

\section{Summary}\label{sec:summary}

Implications of the observed stellar mass--star formation rate correlation
are investigated in the context of current theories for stellar
mass assembly.  The key point, found here and pointed out in previous
studies~\citep{dad07,elb07}, is that the amplitude of the $M_*-$SFR
relation evolves much more rapidly since $z\sim 2$ in observations that
in current galaxy formation models.  It is shown here that this is true
of both hydrodynamic simulations and semi-analytic models, and is broadly
independent of feedback parameters.  In contrast, the slope and scatter
of the observed $M_*-$SFR relation are in good agreement with models.

The tight $M_*-$SFR relation with a slope near unity predicted in models
is a generic result owing to the dominance of cold mode accretion,
particularly in early galaxies, which produces rapid, smooth and
relatively steady infall.  The slow amplitude evolution arises because
star formation starts at $z\ga 6$ for moderately massive star-forming
galaxies and continues at a fairly constant or mildly declining level
to low-$z$.  The large discrepancy in amplitude evolution when compared to
observations from $z\sim 2\rightarrow 0$ hints at some underlying problem
either with the models or with the interpretation of data.  A convenient
parameterization of the problem is through the star formation activity
parameter, $\asf\equiv (M_*/{\rm SFR})/(t_H-1\;{\rm Gyr})$, where $t_H$
is the Hubble time.  A low value corresponds to a starbursting system,
a high value to a passive system, and a value near unity to system forming
stars constantly for nearly a Hubble time.
In models, $\asf$ remains constant around
unity from $z=0-4$, whereas in observations it rises steadily from $\la
0.2$ at $z\sim 2$ to close to unity at $z\sim 0$.

Several ad hoc modifications to the theoretical picture of stellar mass
assembly are considered in order to match $\asf$ evolution, but each
one is found to be in conflict with other observations of high-redshift
galaxies.  Bursts seem unlikely given the low scatter in $M_*-$SFR.
Delaying galaxy formation to match $\asf$ results in such a low redshift
for the onset of star formation, it is in conflict with observations
of star forming galaxies at earlier epochs.  Hiding a large population
of galaxies as passive runs into difficulty when compared to direct
observations of the passive galaxy fraction in mass-selected samples at
$z\sim 2$.  Having an exponentially growing phase of star formation or
``staged" galaxy formation are perhaps the most plausible solutions
from an observational viewpoint, but it is difficult to understand
theoretically how such scenarios can arise within hierarchical structure
formation.  Hence if observed star-forming galaxies are quiescently
forming stars and represent the majority of galaxies at $z\sim 2$, as
other observations seem to suggest, then it is not easy to accomodate
the low $\asf$ inferred from the $M_*-$SFR amplitude.

Systematic uncertainties in $M_*$ or SFR determinations could bias
results progressively more at high-$z$ in order to mimic evolution in
$\asf$.  Various currently debated sources are considered, such as the
contribution to near-IR light from TP-AGB stars, extinction corrections,
assumptions about star formation history, AGN contamination, and PAH
emission calibration.  It may be possible to concoct scenarios whereby
several of these effects combine to mimic $\asf$ evolution, but there
are no suggestions from local observations that such scenarios are to
be expected.  Hence if systematic effects are to explain the low $\asf$
at high-$z$, it would imply significant and unexpected changes in tracers
of star formation and stellar mass between now and high redshifts.

A solution is proposed that the stellar initial mass function becomes
increasingly bottom-light to higher redshifts.  Several lines of arguments
are presented that vaguely or circumstantially favor such evolution,
though no smoking gun signatures are currently known.  A simple model
of IMF evolution is constructed, based on the ansatz by \citet{lar98}
that the minimum temperature of molecular clouds is reflected in the
characteristic mass of star formation ($\hat{M}_{\rm IMF}$) where the
mass contribution per logarithmic mass bin is maximized.  The minimum
temperature may increase with redshift owing to a hotter cosmic microwave
background, more vigorous star formation activity, or lower metallicities
within early galactic ISMs.  In order to reconcile the observed $\asf$
evolution with theoretical expectations of no evolution, an evolving
IMF of the form
\begin{eqnarray*}
\frac{dN}{d\log{M}} &\propto& M^{-0.3}\;\; {\rm for}\; M<\hat{M}_{\rm IMF},\\
&\propto& M^{-1.3}\;\; {\rm for}\; M>\hat{M}_{\rm IMF};\\
\hat{M}_{\rm IMF}&=&0.5 (1+z)^{2} M_\odot
\end{eqnarray*}
is proposed.  The exponent of $\hat{M}_{\rm IMF}$ evolution is constrained
by requiring no $\asf$ evolution, through careful modeling with the
PEGASE.2 population synthesis code.  While the exact form of IMF evolution
is not well constrained by present observations, what is required is that
the IMF has progressively more high-mass stars compared to low-mass at
earlier epochs, and that this IMF applies to the majority of star forming
galaxies at any epoch.  It is worth noting that this evolving IMF is only
constrained out to $z\sim 2$ from the $M_*-$SFR relation, though it 
yields predictions that are consistent with other observations out to
$z\sim 4$.  Extrapolating such evolution to higher redshifts
is dangerous, since no observational constraints exist and the precise
cause of IMF evolution is not understood.

Implications of such an evolving IMF are investigated.  By leaving the
high-mass end of the IMF unchanged, recent successes in understanding
the connections between high-mass star formation, feedback, and metal
enrichment are broadly preserved.  The cosmic stellar mass accumulation
rate would be altered compared to what is inferred from cosmic star
formation history measurements using a standard IMF.  It is shown that an
evolving IMF is at face value in better agreement with direct measures of
cosmic stellar mass assembly~\citep{per07,els07,wil07}.
Furthermore, the evolving IMF goes towards relieving the generic tension
between present-day fossil light measures versus observations of the
cosmic star formation history.  In particular, the paunchy IMF favored by
\citet{far07} in order to reconcile the observed cosmic star formation
history, present-day $K$-band luminosity density, and extragalactic
background light constraints, is qualitatively similar to the cumulative
IMF of all stars formed by today in the evolving IMF case.  Individually,
each argument has sufficient uncertainties to cast doubt on whether 
a radical solution such as an evolving IMF is necessary.  But taken together,
the $M_*-$SFR relation adds to a growing body of circumstantial evidence
that the ratio of high-mass to low-mass stars formed is higher at
earlier epochs.

It is by no means clear that IMF variations are the only viable solution
to the $M_*-$SFR dilemma.  The claim here is only that an evolving IMF is
an equally (un)likely solution as invoking unknown systematic effects or
carefully crafted star formation histories in order to explain $M_*-$SFR
evolution.  It is hoped that this work will spur further efforts, both
observational and theoretical, to investigate this important issue.

Future plans include performing a more careful analysis of the evolving
IMF in terms of observable quantities, in order to accurately
quantify the impact of such IMF evolution on the interpretation of UV,
near-IR, and mid-IR light.  Also, it is feasible to incorporate such an
evolving IMF directly into simulation runs to properly account
for gas recycling in such a scenario.  Finally, including some feedback
mechanism to truncate star formation in massive systems may impact the
$M_*-$SFR relation in some way, so this will be incorporated into the
hydro simulations.

Observationally, pushing SFR and $M_*$ determinations to higher redshifts
is key; a continued drop in $\asf$ to $z\sim 3$ would rapidly solidify
the discrepancies with current models, and would also generate stronger
conflicts with other observations of high-$z$ galaxies.  Assessing the
AGN contribution and extinction uncertainties from high-$z$ systems
is critical for accurately quantifying the light from high-mass star
formation, for instance through the use of more direct star formation
indicators such as Paschen-$\alpha$.  Pushing observations further into
the mid-IR such as to 70$\mu$, past the PAH bands at $z\sim 2$, would
mitigate PAH calibration uncertainties in SFR estimates.  Obtaining a
large sample of spectra for typical high-$z$ star-forming systems
\citep[like cB58;][]{pet00} would more accurately constrain the SED
than broad-band data.  All of these programs push current technological
capabilities to their limits and perhaps beyond, but are being planned
as facilities continue their rapid improvement.

In summary, owing to the robust form of star formation histories in
current galaxy formation models, the $M_*$-SFR relation represents a
key test of our understanding of stellar mass assembly.  Current models
reproduce the observed slope and scatter remarkably well, but broadly fail
this test in terms of amplitude evolution.  Whether this reflects some
fundamental lack of physical insight, or else some missing ingredient
such as an evolving IMF, is an issue whose resolution will have a
significant impact on our understanding of galaxy formation.

 \section*{Acknowledgements}
The author is grateful for valuable discussions with Emanuele Daddi,
Arjun Dey, Mark Dickinson, David Elbaz, Mark Fardal, Kristian Finlator,
Karl Glazebrook, Andrew Hopkins, Du\^san Kere\^s, Janice Lee, Heather
Morrison, Kai Noeske, Casey Papovich, Naveen Reddy, Greg Rudnick,
J. D. Smith, Pieter van Dokkum, David Weinberg, and Steve Wilkins.
The author appreciates stimulus from Neal Katz, who been espousing
a top-heavy IMF at high redshifts for some time.  The author thanks
H.-W. Rix and the Max-Planck Instit\"ut f\"ur Astronomie for their
gracious hospitality while some of this work was being done.  The
simulations used here were run by Ben D. Oppenheimer on Grendel, our
department's Beowulf system at the University of Arizona, and the Xeon
Linux Cluster at the National Center for Supercomputing Applications.
Support for this work, part of the Spitzer Space Telescope Theoretical
Research Program, was provided by NASA through a contract issued by the
Jet Propulsion Laboratory, California Institute of Technology under a
contract with NASA.  Support for this work was also provided by NASA
through grant number HST-AR-10946 from the SPACE TELESCOPE SCIENCE
INSTITUTE, which is operated by AURA, Inc. under NASA contract NAS5-26555.


\begin{thebibliography}{}
\frenchspacing

\bibitem[Bell et al. (2007)]{bel07} Bell, E. F., Zheng, X. Z., Papovich, C., Borch, A., Wolf, C., Meisenheimer, K. 2007, ApJ, 663, 834
\bibitem[Baldry \& Glazebrook (2003)]{bal03} Baldry, I. K. \& Glazebrook, K. 2003, ApJ, 593, 258
\bibitem[Birnboim \& Dekel (2003)]{bir03} Birnboim, Y. \& Dekel, A. 2003, MNRAS, 345, 349
\bibitem[Borch et al. (2006)]{bor06} Borch, A., Meisenheimer, K., Bell, E. F., Rix, H.-W., Wolf, C., Dye, S., Kleinheinrich, M., Kovacs, Z., Wisotzki, L. 2006, A\&A, 453, 869
\bibitem[Bouwens et al. (2007)]{bou07} Bouwens, R. J., Illingworth, G. D., Franz, M., Ford, H. C. 2007, ApJ, in press,  arXiv:0707.2080
\bibitem[Brooks et al. (2007)]{bro07} Brooks, A. M., Governato, F., Booth, C. M., Willman, B., Gardner, J. P., Wadsley, J., Stinson, G., Quinn, T. 2007, ApJL, 655, L17
\bibitem[Bruzual (2007)]{bru07} Bruzual, G. A. 2007, in proc. IAU Symposium No. 241 ``Stellar populations as building blocks of galaxies", eds. A. Vazdekis and R. Peletier, Cambridge: Cambridge University Press, astro-ph/0703052
\bibitem[Bruzual \& Charlot (2003)]{bru03} Bruzual, G. A. \& Charlot, S. 2003, MNRAS, 344, 1000
\bibitem[Calzetti et al. (2000)]{cal00} Calzetti, D., Armus, L., Bohlin, R. C., Kinney, A., Koornneef, J., Storchi-Bergmann, T. 2000, ApJ, 533, 682
\bibitem[Calzetti et al. (2007)]{cal07} Calzetti, D., et al. 2007, ApJ, 666, 870
\bibitem[Chabrier (2003)]{cha03} Chabrier, G. 2003, PASP, 115, 763
\bibitem[Chary (2007)]{cha07} Chary, R.-R. 2007, ApJ, submitted, arXiv:0712.1498 
\bibitem[Cole et al. (2001)]{col01} Cole, S. et al. 2001, MNRAS, 326, 255
\bibitem[Cowie et al. (1996)]{cow96} Cowie, L. L., Songaila, A., Hu, E. M., Cohen, J. G. 1996, AJ, 112, 839
\bibitem[Daddi et al. (2007)]{dad07} Daddi, E., Dickinson, M., Morrison, G., Chary, R., Cimatti, A., Elbaz, D., Frayer, D., Renzini, A., Pope, A., Alexander, D. M., Bauer, F. E., Giavalisco, M., Huynh, M., Kurk, J., Mignoli, M. 2007, ApJ, submitted, arXiv:0705.2831
\bibitem[Daddi et al. (2007b)]{dad07b} Daddi, E., Alexander, D. M., Dickinson, M., Gilli, R., Renzini, A., Elbaz, D., Cimatti, A., Chary, R., Frayer, D., Bauer, F. E., Brandt, W. N., Giavalisco, M., Grogin, N. A., Huynh, M., Kurk, J., Mignoli, M., Morrison, G., Pope, A., Ravindranath, S. 2007, ApJ, accepted, arXiv:0705.2832
\bibitem[Dale et al. (2006)]{dal06} Dale, D. A. et al. 2006, ApJ, 646, 161
%\bibitem[Dav\'e et al. (1999)]{dav99} Dav\'e, R., Hernquist, L., Katz, N., \& Weinberg, D. H. 1999, ApJ, 511, 521
%\bibitem[Dav\'e et al. (2001)]{dav01} Dav\'e, R., Cen, R., Ostriker, J. P., Bryan, G. L., Hernquist, L., Katz, N., Weinberg, D. H., Norman, M. L., \& O'Shea, B. 2001, ApJ, 552, 473
\bibitem[Dav\'e, Finlator \& Oppenheimer (2006)]{dav06} Dav\'e, R., Finlator, K., \& Oppenheimer, B. D. 2006, MNRAS, 370, 273
\bibitem[Dav\'e \& Oppenheimer (2007)]{dav07} Dav\'e, R. \& Oppenheimer, B. D. 2007, MNRAS, 374, 427
\bibitem[De Lucia \& Blaizot (2007)]{del07} De Lucia, G. \& Blaizot, J. 2007, MNRAS, 375, 2
\bibitem[Drory et al. (2005)]{dro05} Drory, N., Salvato, M., Gabasch, A., Bender, R., Hopp, U., Feulner, G., Pannella, M. 2005, ApJL, 619, L131
\bibitem[Elbaz et al. (2007)]{elb07} Elbaz, D., Daddi, E., Le Borgne, D., Dickinson, M., Alexander, D. M., Chary, R.-R., Starck, J.-L., Brandt, W. N., Kitzbichler, M., MacDonald, E., Nonino, M., Popesso, P., Stern, D., Vanzella, E. 2007, A\&A, 468, 33
\bibitem[Elsner et al. (2007)]{els07} Elsner, F., Feulner, G., Hopp, U. 2007, A\&A, in press, arXiv:0711.0384
\bibitem[Erb et al. (2006a)]{erb06} Erb, D. K., Shapley, A. E., Pettini, M., Steidel, C. C., Reddy, N. A, \& Adelberger, K. L. 2006a, ApJ, 644, 813
\bibitem[Erb et al. (2006b)]{erb06b} Erb, D. K., Steidel, C. C., Shapley, A. E., Pettini, M., Reddy, N. A., Adelberger, K. L. 2006b, ApJ, 646, 107
\bibitem[Erb et al. (2006c)]{erb06c} Erb, D. K., Steidel, C. C., Shapley, A. E., Pettini, M., Reddy, N. A., Adelberger, K. L. 2006c, ApJ, 647, 128
\bibitem[Fardal et al. (2007)]{far07} Fardal, M. A., Katz, N., Weinberg, D. H., Dav\'e, R., Katz, N. 2007, MNRAS, 379, 985
\bibitem[Ferguson et al. (2002)]{fer02} Ferguson, H. C., Dickinson, M., Papovich, C. 2002, ApJL, 569, L65
\bibitem[Figer (2005)]{fig05} Figer, D. F. 2005, in ``The Formation and Evolution of Massive Young Star Clusters", ASP conf. series, Vol. 322. eds. H.J.G.L.M. Lamers, L.J. Smith, A. Nota, San Francisco: ASP, p.49
\bibitem[Finlator et al. (2006)]{fin06} Finlator, K., Dav\'e, R., Papovich, C., \& Hernquist, L. 2006, ApJ, 639, 672
\bibitem[Finlator, Dav\'e \& Oppenheimer (2007)]{fin07} Finlator, K., Dav\'e, Oppenheimer, B. D. 2007, MNRAS, 376, 1861
\bibitem[Finlator \& Dav\'e (2007)]{fin07b} Finlator, K. \& Dav\'e 2007, MNRAS, submitted, arXiv:0704.3100
\bibitem[Fioc \& Rocca-Volmerange (1997)]{fio97} Fioc, M., Rocca-Volmerange, B. 1997, A\&A, 326, 950
\bibitem[Fontana et al. (2004)]{fon04} 	Fontana, A., Pozzetti, L., Donnarumma, I., Renzini, A., Cimatti, A., Zamorani, G., Menci, N., Daddi, E., Giallongo, E., Mignoli, M., Perna, C., Salimbeni, S., Saracco, P., Broadhurst, T., Cristiani, S., D'Odorico, S., Gilmozzi, R. 2004, A\&A, 424, 23
\bibitem[Fontana et al. (2006)]{fon06} 	Fontana, A. et al. 2006, A\&A, 459, 745
\bibitem[Guo \& White (2007)]{guo07} Guo, Q. \& White, S. D. M. 2007, arXiv:0708.1814
\bibitem[Heavens et al. (2004)]{hea04} Heavens, A., Panter, B., Jimenez, R., Dunlop, J. 2004, Nature, 428, 625
\bibitem[Hopkins (2004)]{hop04} Hopkins, A. M. 2004, ApJ, 615, 209
\bibitem[Hopkins \& Beacom (2006)]{hop06} Hopkins, A. M. \& Beacom, J. F. 2006, ApJ, 651, 142
\bibitem[Hopkins et al. (2007)]{hop07} Hopkins, P. F., Cox, T. J., Kere\^s, D., Hernquist, L. 2007, ApJ, submitted, arXiv:0706.1246
\bibitem[Jappsen et al. (2005)]{jap05} Jappsen, A.-K., Klessen, R. S., Larson, R. B., Li, Y., Mac Low, M.-M. 2005, A\&A, 435, 611
\bibitem[Katz et al. (2003)]{kat03} Katz, N., Kere\^s, D., Dav\'e, R., Weinberg, D. H. 2003, in ``IGM/Galaxy Connection- The Distribution of Baryons at $z=0$", ASSL conf. series vol. 281, eds. J. L. Rosenberg \& M. E. Putman, Kluwer, Dodrecht, p.185
\bibitem[Keres et al. (2005)]{ker05} Keres, D., Katz, N., Weinberg, D. H., \& Dav\'e, R. 2005, MNRAS, 363, 2
\bibitem[Kennicutt (1998)]{ken98} Kennicutt, R. C. 1998, ApJ, 498, 541
\bibitem[Kitzbichler \& White (2007)]{kit07} Kitzbichler, M. G., White, S. D. M. 2007, MNRAS, 376, 2
\bibitem[Kobayashi, Springel \& White (2007)]{kob07} Kobayashi, C., Springel, V., White, S. D. M. 2007, MNRAS, 376, 1465
\bibitem[Kolatt et al. (1999)]{kol99} Kolatt, T. S., Bullock, J. S., Somerville, R. S., Sigad, Y., Jonsson, P., Kravtsov, A. V., Klypin, A. A., Primack, J. R., Faber, S. M., Dekel, A. 1999, 523, L109
\bibitem[Kroupa (2001)]{kro01} Kroupa, P. 2001, MNRAS, 322, 231
\bibitem[Kroupa (2007)]{kro07} Kroupa, P. 2007, in proc. ``Resolved Stellar Populations", eds. D. Valls-Gabaud and M. Chavez, ASP Conf. Ser. (in press), astro-ph/0703124
\bibitem[Kroupa, Tout, \& Gilmore (1993)]{kro93} Kroupa, P., Tout, C. A., Gilmore, G. 1993, MNRAS, 262, 545
%\bibitem[Labb\'e et al. (2007)]{lab07} Labb\'e, I., Franx, M., Rudnick, G., F\"orster Schreiber, N. M., van Dokkum, P. G., Moorwood, A., Rix, H.-W., Röttgering, H., Trujillo, I., van der Werf, P. 2007, ApJ, 665, 944
\bibitem[Lacey et al. (2007)]{lac07} Lacey, C. G., Baugh, C. M., Frenk, C. S., Silva, L., Granato, G. L., Bressan, A. 2007, MNRAS, submitted, arXiv:0704.1562
\bibitem[Larson (1985)]{lar85} Larson, R. B. 1985, MNRAS, 214, 379
\bibitem[Larson (1998)]{lar98} Larson, R. B. 1998, MNRAS, 301, 569
\bibitem[Larson (2005)]{lar05} Larson, R. B. 2005, MNRAS, 359, 211
\bibitem[Li et al. (2007)]{li07} Li, Y., Mo, H. J., van den Bosch, F. C., Lin, W. P. 2007, MNRAS, 379, 689
\bibitem[Lucatello et al. (2005)]{luc05} Lucatello, S., Gratton, R. G., Beers, T. C., Carretta, E. 2005, ApJ, 625, 833
%\bibitem[Madau et al. (1996)]{mad96} Madau, P., Ferguson, H. C., Dickinson, M. E., Giavalisco, M., Steidel, C. C., Fruchter, A. 1996, MNRAS, 283, 1388
\bibitem[Madau, Pozzetti, \& Dickinson (1998)]{mad98} Madau, P., Pozzetti, L., Dickinson, M. 1998, ApJ, 498, 106
\bibitem[Maraston et al. (2006)]{mar06} Maraston, C., Daddi, E., Renzini, A., Cimatti, A., Dickinson, M., Papovich, C., Pasquali, A., Pirzkal, N. 2006, ApJ, 652, 85
\bibitem[Miller \& Scalo (1979)]{mil79} Miller, G. E. \& Scalo, J. M. 1979, ApJS, 41, 513
\bibitem[Mo, Mao, \& White(1998)]{mo98} Mo, H. J., Mao, S., White, S. D. M. 1998, MNRAS, 295, 319
\bibitem[Murali et al. (2001)]{mur01} Murali, C., Katz, N., Hernquist, L., Weinberg, D. H., Dav\'e, R. 2002, ApJ, 571, 1
\bibitem[Neistein, van den Bosch \& Dekel (2006)]{nei06} Neistein, E., van den Bosch, F. C., Dekel, A. 2006, MNRAS, 372, 933
\bibitem[Noeske et al. (2007a)]{noe07a} Noeske, K. G., et al. 2007a, ApJL, 660, L43
\bibitem[Noeske et al. (2007b)]{noe07b} Noeske, K. G., et al. 2007b, ApJL, 660, L47
\bibitem[Oppenheimer \& Dav\'e (2006)]{opp06} Oppenheimer, B. D. \& Dav\'e, R. 2006, MNRAS, 373, 1265
\bibitem[Oppenheimer \& Dav\'e (2007)]{opp07} Oppenheimer, B. D. \& Dav\'e, R. 2007, in preparation
\bibitem[Papovich, Dickinson \& Ferguson (2001)]{pap01} Papovich, C., Dickinson, M., Ferguson, H. C. 2001, ApJ, 559, 620
\bibitem[Papovich et al. (2006)]{pap06} Papovich, C. et al. 2006, ApJ, 640, 92
\bibitem[Perez-Gonzalez et al. (2007)]{per07} Perez-Gonzalez, P. G., Rieke, G. H., Villar, V., Barro, G., Blaylock, M., Egami, E., Gallego, J., Gil de Paz, A., Pascual, S., Zamorano, J., Donley, J. L. 2007, ApJ, accepted, arXiv:0709.1354
\bibitem[Pettini et al. (2000)]{pet00} Pettini, M., Steidel, C. C., Adelberger, K. L., Dickinson, M., Giavalisco, M. 2000, ApJ, 528, 96
%\bibitem[Pettini et al. (2001)]{pet01} Pettini, M., Shapley, A. E., Steidel, C. C., Cuby, J.-G., Dickinson, M., Moorwood, A. F. M., Adelberger, K. L., Giavalisco, M. 2001, ApJ, 554, 981
\bibitem[Portinari et al. (2004)]{por04} Portinari, L., Moretti, A., Chiosi, C., Sommer-Larsen, J. 2004, ApJ, 604, 579
\bibitem[Reddy et al. (2005)]{red05} Reddy, N. A., Erb, D. K., Steidel, C. C., Shapley, A. E., Adelberger, K. L., Pettini, M. 2005, ApJ, 633, 748
\bibitem[Reddy et al. (2007)]{red07} Reddy, N. A., Steidel, C. C., Pettini, M., Adelberger, K. L., Shapley, A. E., Erb, D. K., Dickinson, M. 2007, ApJ, in press
\bibitem[Renzini (2005)]{ren05} Renzini, A. 2005, in ``The Initial Mass Function 50 years later," eds. E. Corbelli and F. Palle, ASSL, Spinger conf. ser. v.327, p.221
\bibitem[Rieke et al. (1980)]{rie80} Rieke, G. H., Lebofsky, M. J., Thompson, R. I., Low, F. J., Tokunaga, A. T. 1980, ApJ, 238, 24
\bibitem[Rieke et al. (1993)]{rie93} Rieke, G. H., Loken, K., Rieke, M. J., Tamblyn, P. 1993, ApJ, 412, 99
\bibitem[Rudnick et al. (2006)]{rud06} Rudnick, G. et al. 2006, ApJ, 650, 624
\bibitem[Salim et al. (2007)]{sal07} Salim, S. et al. 2007, ApJ, in press, arXiv:0704.3611
\bibitem[Savaglio et al. (2005)]{sav05}	Savaglio, S., et al. 2005, ApJ, 635, 260
\bibitem[Schechter (1976)]{sch76} Schechter, P. 1976, ApJ, 203, 297
\bibitem[Scott et al. (2002)]{sco02} Scott, J. E., Bechtold, J., Morita, M., Dobrzycki, A., Kulkarni, V. 2002, ApJ, 571, 665
\bibitem[Scoville et al. (2007)]{sco07} Scoville, N. et al. 2007, ApJS, 172, 150
%\bibitem[Shapley et al. (2004)]{sha04} Shapley, A. E., Erb, D. K., Pettini, M., Steidel, C. C., Adelberger, K. L. 2004, ApJ, 612, 108
\bibitem[Shapley et al. (2005)]{sha05} Shapley, A. E., Steidel, C. C., Erb, D. K., Reddy, N. A., Adelberger, K. L., Pettini, M., Barmby, P., Huang, J. 2005, ApJ, 626, 698
\bibitem[Sirianni et al. (2000)]{sir00} Sirianni, M., Nota, A., Leitherer, C., De Marchi, G., Clampin, M. 2000, ApJ, 533, 203
\bibitem[Smith et al. (2007)]{smi07} Smith, J.-D. T., et al. 2007, ApJ, 656, 770
%\bibitem[Somerville, Primack \& Faber (2001)]{som01} Somerville, R. S., Primack, J. R., Faber, S. M. 2001, MNRAS, 320, 504
\bibitem[Springel \& Hernquist (2003)]{spr03} Springel, V. \& Hernquist, L. 2003, MNRAS, 339, 312
\bibitem[Springel (2005)]{spr05} Springel, V. 2005, MNRAS, 364, 1105
%\bibitem[Tremaine et al. (2002)]{tre02} Tremaine, S. et al. 2002, ApJ, 574, 740
\bibitem[Tumlinson (2007)]{tum07} Tumlinson, J. 2007, ApJ, 664, L63
\bibitem[Van Dokkum \& van der Marel (2007)]{van07} van Dokkum, P. G. \& van der Marel, R. P. 2007, ApJ, 655, 30
\bibitem[Van Dokkum (2007)]{van07b} van Dokkum, P. G. 2007, ApJ, accepted, arXiv:0710.0875
\bibitem[Weidner \& Kroupa (2006)]{wei06} Weidner, C. \& Kroupa, P. 2006, MNRAS, 365, 1333
%\bibitem[Weinberg et al.(2004)]{wei04} Weinberg, D.~H., Dav{\'e}, R., Katz, N., \& Hernquist, L.\ 2004, ApJ, 601, 1 
\bibitem[Wilkins, Trentham \& Hopkins (2007)]{wil07} Wilkins, S. M., Trentham, N., Hopkins, A. M. 2007, MNRAS, accepted, arXiv:0801.1594


\end{thebibliography}
\end{document}